\documentclass[11pt]{article}
\usepackage{times}
\usepackage{mathptm}
\usepackage{graphicx}
\usepackage{anysize}
\usepackage{latexsym}
\usepackage{amsfonts}
\usepackage{color}
\marginsize{2cm}{2cm}{1cm}{3cm}

\newenvironment{definition}
    {
    \smallskip
    \refstepcounter{theorem}
    \noindent
    {\bf Definition \arabic{section}.\arabic{theorem}} \ \ }
    {\hspace*{\fill}{\ }
    \smallskip}

\newenvironment{app_theorem}
    {
    \smallskip
    \refstepcounter{theorem}
    \noindent
    {\bf Theorem A.\arabic{theorem}} \ \ }
    {\hspace*{\fill}{\ }
    \smallskip}

\newenvironment{app_lemma}
    {
    \smallskip
    \refstepcounter{theorem}
    \noindent
    {\bf Lemma A.\arabic{theorem}} \ \ }
    {\hspace*{\fill}{\ }
    \smallskip}

\newtheorem{theorem}{Theorem}[section]

\newtheorem{lemma}[theorem]{Lemma}
\newtheorem{corollary}[theorem]{Corollary}

\begin{document}
\pagestyle{plain}

\title{Improving Gate-Level Simulation of Quantum
Circuits\thanks{Earlier results of this work were reported at ASPDAC
'03 \cite{ViamontesEtAl002}. New material includes significantly
better experimental results and a description of a class of
matrices and vectors which can be manipulated in polynomial time and
memory using QuIDDPro.}}

\author{George F. Viamontes, Igor L. Markov, and John P. Hayes \\
  \{gviamont, imarkov, jhayes\}@eecs.umich.edu \\
  The University of Michigan, Advanced Computer Architecture Laboratory \\
  Ann Arbor, MI~~48109-2122, USA}

\maketitle

\abstract {Simulating quantum computation on a classical computer is a
  difficult problem.  The matrices representing quantum gates, and
  the vectors modeling qubit states grow exponentially with an increase
  in the number of qubits.  However, by using a novel data structure
  called the Quantum Information Decision Diagram (QuIDD) that
  exploits the structure of quantum operators, a useful subset of
  operator matrices and state vectors can be represented in a form that grows
  polynomially with the number of qubits. This
  subset contains, but is not limited to, any equal superposition of
  $n$ qubits, any computational basis state, $n$-qubit Pauli matrices,
  and $n$-qubit Hadamard matrices. It does not, however, contain
  the discrete Fourier
  transform (employed in Shor's algorithm) and some oracles used in Grover's
  algorithm. We first introduce and motivate decision diagrams and
  QuIDDs. We then
  analyze the runtime and memory complexity of QuIDD operations. Finally,
  we empirically validate QuIDD-based simulation by means of a
  general-purpose quantum computing simulator QuIDDPro implemented in
  C++. We simulate various instances of Grover's
  algorithm with QuIDDPro, and the results demonstrate that QuIDDs
  asymptotically outperform all other known simulation techniques.
  Our simulations also show that well-known worst-case
  instances of classical searching can be circumvented in many specific
  cases by data compression techniques.

\newpage

\section{Introduction}

Richard Feynman observed in the 1980s that simulating quantum
mechanical processes on a standard {\em classical} computer seems to
require super-polynomial memory and time \cite{Hey99}. For instance, a
complex vector of size $2^n$ is needed to represent all the
information in $n$ quantum states, and square matrices of size
$2^{2n}$ are needed to model (simulate) the time evolution of the
states \cite{Nielsen2000}. Consequently, Feynman proposed {\em quantum
computing} which uses the quantum mechanical states themselves to
simulate quantum processes. The key idea is to replace bits with
quantum states called {\em qubits} as the fundamental units of
information. A quantum computer can operate directly on exponentially
more data than a classical computer with a similar number of
operations and information units. Thus in addressing the problem of
simulating quantum mechanical processes more efficiently, Feynman
discovered a new computing model that can outperform classical
computation in certain cases.

Software simulation has long been an invaluable tool for the design
and testing of digital circuits. This problem too was once thought to
be computationally intractable. Early simulation and synthesis
techniques for $n$-bit circuits often required $O(2^n)$ runtime and
memory, with the worst-case complexity being fairly typical. Later
algorithmic advancements brought about the ability to perform circuit
simulation much more efficiently in practical cases. One such advance
was the development of a data structure called the Reduced Ordered
Binary Decision Diagram (ROBDD) \cite{bryant}, which can greatly
compress the Boolean description of digital circuits and allow direct
manipulation of the compressed form. Software simulation may also play
a vital role in the development of quantum hardware by enabling the
modeling and analysis of large-scale designs that cannot be
implemented physically with current technology. Unfortunately,
straightforward simulation of quantum designs by classical computers
executing standard linear-algebraic routines requires $O(2^n)$ time
and memory \cite{Hey99, Nielsen2000}. However, just as ROBDDs and
other innovations have made the simulation of very large classical
computers tractable, new algorithmic techniques can allow the
efficient simulation of quantum computers.

The goal of the work reported here is to develop a practical software
means of simulating quantum computers efficiently on classical
computers. We propose a new data structure called the {\em Quantum
Information Decision Diagram} (QuIDD) which is based on decision
diagram concepts that are well-known in the context of simulating
classical computer hardware \cite{Clarke96, Bahar97, bryant}. As we
demonstrate, QuIDDs allow simulations of $n$-qubit systems to achieve
run-time and memory complexities that range from $O(1)$ to $O(2^n)$,
and the worst case is not typical. In the important case of Grover's
quantum search algorithm \cite{Grover97}, we show that our QuIDD-based
simulator outperforms all other known simulation techniques.

The paper is organized as follows. Section 2 outlines previous work on decision
diagrams and the modeling of quantum computation on classical computers. In
Section 3 we present our QuIDD data structure. Section 4 analyzes the runtime
and memory complexity of QuIDD operations, while Section 5 describes some
experimental results using QuIDDs. Finally, in Section 6 we present our
conclusions and ideas for future work.

\section{Background}
\label{sec:background}

This section first presents the basic concepts of decision
diagrams, assuming only a rudimentary knowledge of computational
complexity and graph theory. It then reviews previous research on
simulating quantum mechanical matrix operations.

\subsection{Binary Decision Diagrams}
\label{sec:bdd}

The  binary decision diagram (BDD) was introduced by Lee in 1959
\cite{Lee59} in the context of classical logic circuit design.
This data structure represents a Boolean function
$f(x_1,x_2,...,x_n)$ by a directed acyclic graph (DAG); see Figure
\ref{fig:bdd}. By convention, the top node of a BDD is labeled
with the name of the function $f$ represented by the BDD. Each
variable $x_i$ of $f$  is associated with one or more nodes with
two outgoing edges labeled {\em then} (solid line) and {\em else}
(dashed line). The {\em then} edge of node $x_i$ denotes an
assignment of logic $1$ to the $x_i$, while the {\em else} edge
represents an assignment of logic $0$. These nodes are called {\em
internal} nodes and  are labeled by the corresponding  variable
$x_i$. The edges of the BDD point downward, implying a top-down
assignment of values to the Boolean variables depicted by the
internal nodes.

\begin{figure}[htb]
  \begin{center}
    \begin{tabular}{c|c|c|c}
      \parbox{3cm}{
    \begin{center}
      \vspace{-30mm}
      {\large $f=x_0 \cdot x_1 + x_1$ }
    \end{center}
      }
      &
      \includegraphics[width=5cm]{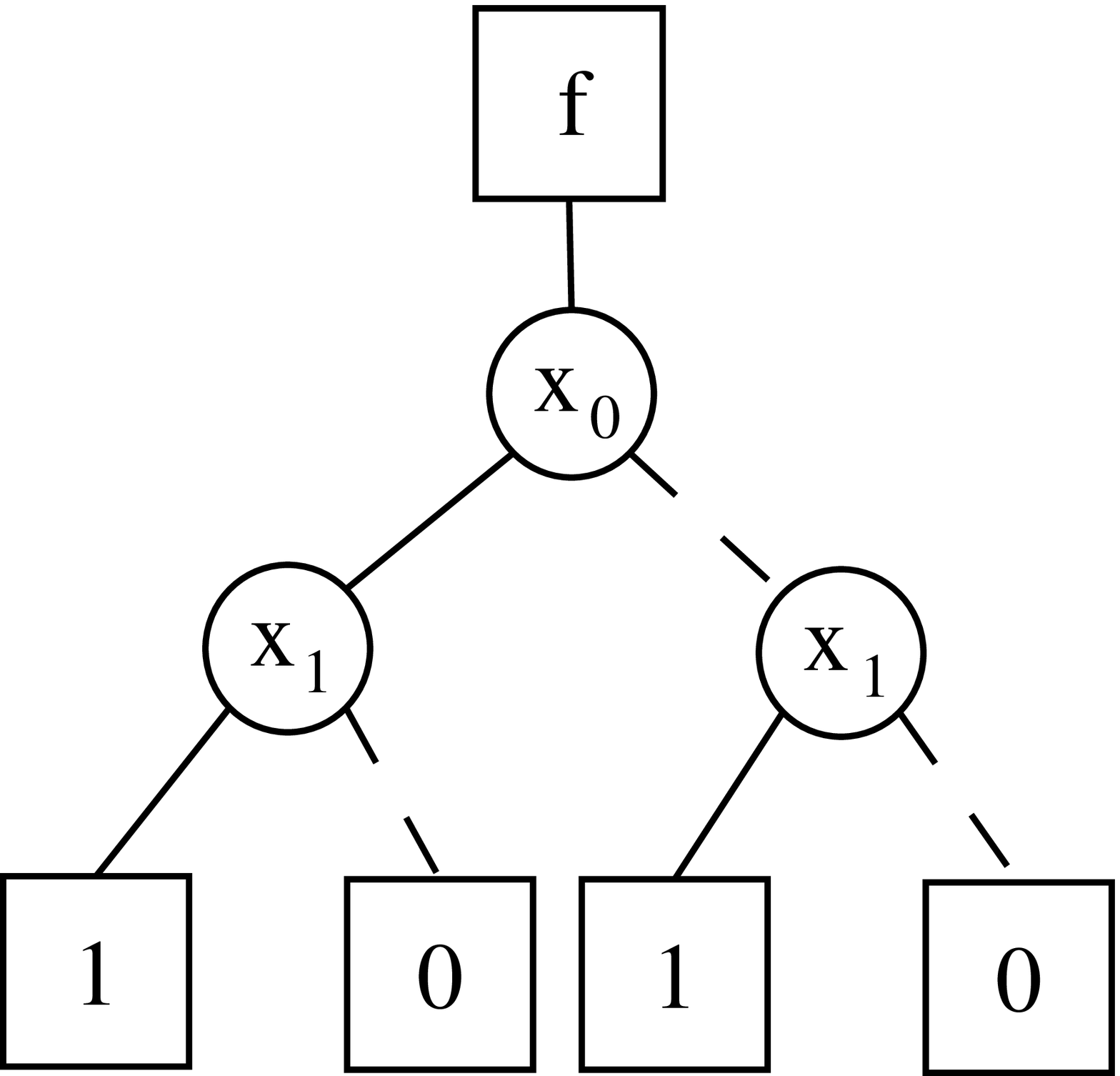}
      &
      \includegraphics[width=2.4cm]{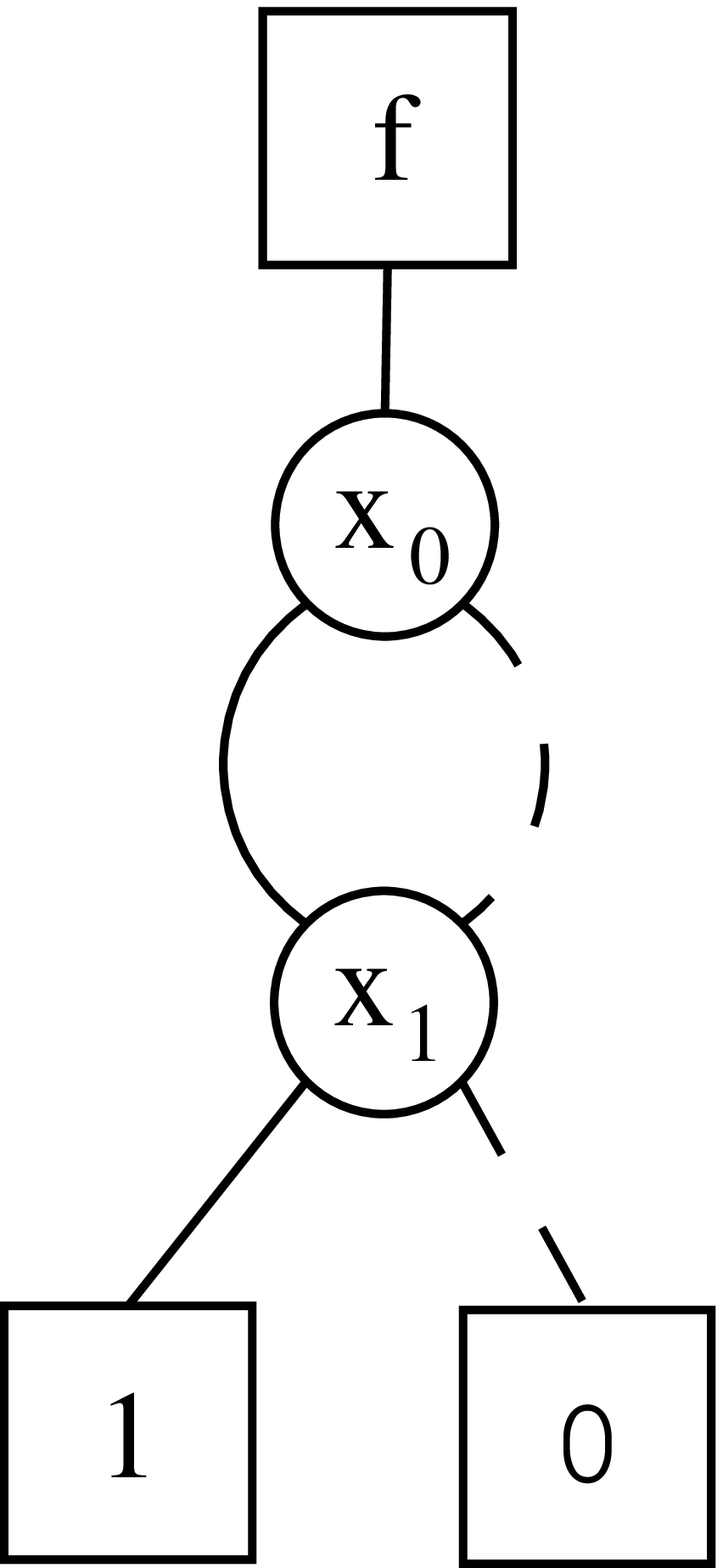}
      &
      \includegraphics[width=2.4cm]{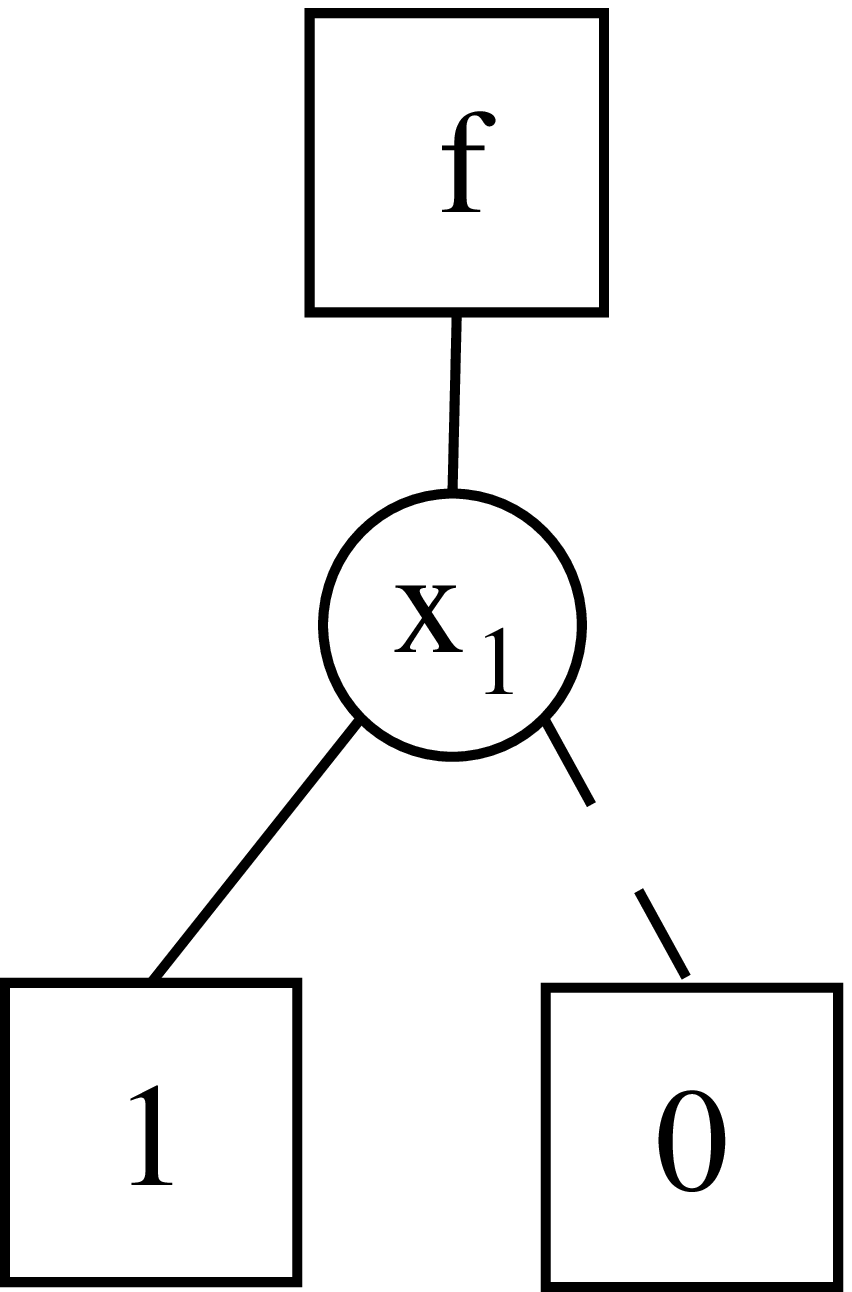}
      \\
      (a) & (b) & (c) & (d)
    \end{tabular}
    \parbox{14cm}{\caption{\label{fig:bdd} (a) A logic function, (b)
    its BDD representation, (c) its BDD representation after applying
    the first reduction rule, and (d) its ROBDD representation.}}
  \end{center}
\end{figure}

At the bottom of the BDD are {\em terminal} nodes containing the
logic values $1$ or $0$. They denote the output value of the
function $f$  for a given assignment of its variables. Each path
through the BDD from top to bottom represents a specific
assignment of 0-1 values to the variables  $x_1,x_2,...,x_n$ of
$f$, and ends with the corresponding output value
$f(x_1,x_2,...,x_n)$.

The original BDD data structure conceived by Lee has exponential
worst-case memory complexity $\Theta (2^n)$, where $n$ is the
number of Boolean variables in a given logic function. Moreover,
exponential memory and runtime are required in many practical
cases, making this data structure impractical for simulation of
large logic circuits. To address this limitation, Bryant developed
the Reduced Ordered BDD (ROBDD) \cite{bryant}, where all variables
are ordered, and decisions are made in that order. A key advantage
of the ROBDD is  that variable-ordering facilitates an efficient
implementation of reduction rules that automatically eliminate
redundancy from the basic BDD representation and may be summarized
as follows:

\begin{enumerate}

  {\item There are no nodes $v$ and $v'$ such that the subgraphs
  rooted at $v$ and $v'$ are isomorphic}

  {\item There are no internal nodes with {\em then} and {\em else}
  edges that both point to the same node}

\end{enumerate}

An example of how the rules transform a BDD into an ROBDD is shown
in Figure  \ref{fig:bdd}. The subgraphs rooted at the
$x_1$ nodes in Figure \ref{fig:bdd}b are isomorphic. By applying
the first reduction rule, the BDD in Figure \ref{fig:bdd}b is
converted into the BDD in Figure \ref{fig:bdd}c. Notice that in
this new BDD, the {\em then} and {\em else} edges of the $x_0$
node now point to the same node. Applying the second reduction
rule eliminates the $x_0$ node, producing the ROBDD in Figure
\ref{fig:bdd}d. Intuitively it makes sense to eliminate the $x_0$
node since the output of the original function is determined
solely by the value of $x_1$. An important aspect of redundancy
elimination is the sensitivity of ROBDD size to the variable
ordering. Finding the optimal variable ordering is an
$NP$-complete problem, but efficient ordering heuristics
have been developed for specific applications. Moreover, it turns
out that many practical logic functions have ROBDD representations
that are polynomial (or even linear) in the number of input
variables \cite{bryant}. Consequently, ROBDDs have become
indispensable tools in the design and simulation of classical
logic circuits.

\subsection{BDD Operations}
\label{sec:bdd_ops}


Even though the ROBDD is often quite compact, efficient algorithms are
necessary to make it practical for circuit simulation. Thus, in
addition to the foregoing reduction rules, Bryant introduced a variety
of ROBDD operations whose complexities are bounded by the size of the
ROBDDs being manipulated \cite{bryant}. Of central importance is the
$Apply$ operation, which performs a binary operation with two ROBDDs,
producing a third ROBDD as the result. It can be used, for example, to
compute the logical $AND$ of two functions. $Apply$ is implemented by
a recursive traversal of the two ROBDD operands. For each pair of
nodes visited during the traversal, an internal node is added to the
resultant ROBDD using the three rules depicted in Figure
\ref{fig:apply_rules}. To understand the rules, some notation must be
introduced. Let $v_f$ denote an arbitrary node in an ROBDD $f$.  If
$v_f$ is an internal node, $Var(v_f)$ is the Boolean variable
represented by $v_f$, $T(v_f)$ is the node reached when traversing the
{\em then} edge of $v_f$, and $E(v_f)$ is the node reached when
traversing the {\em else} edge of $v_f$.

\begin{figure}[htb]
  \begin{center}
    \begin{tabular}{c|c|c}
      \includegraphics[width=4.7cm]{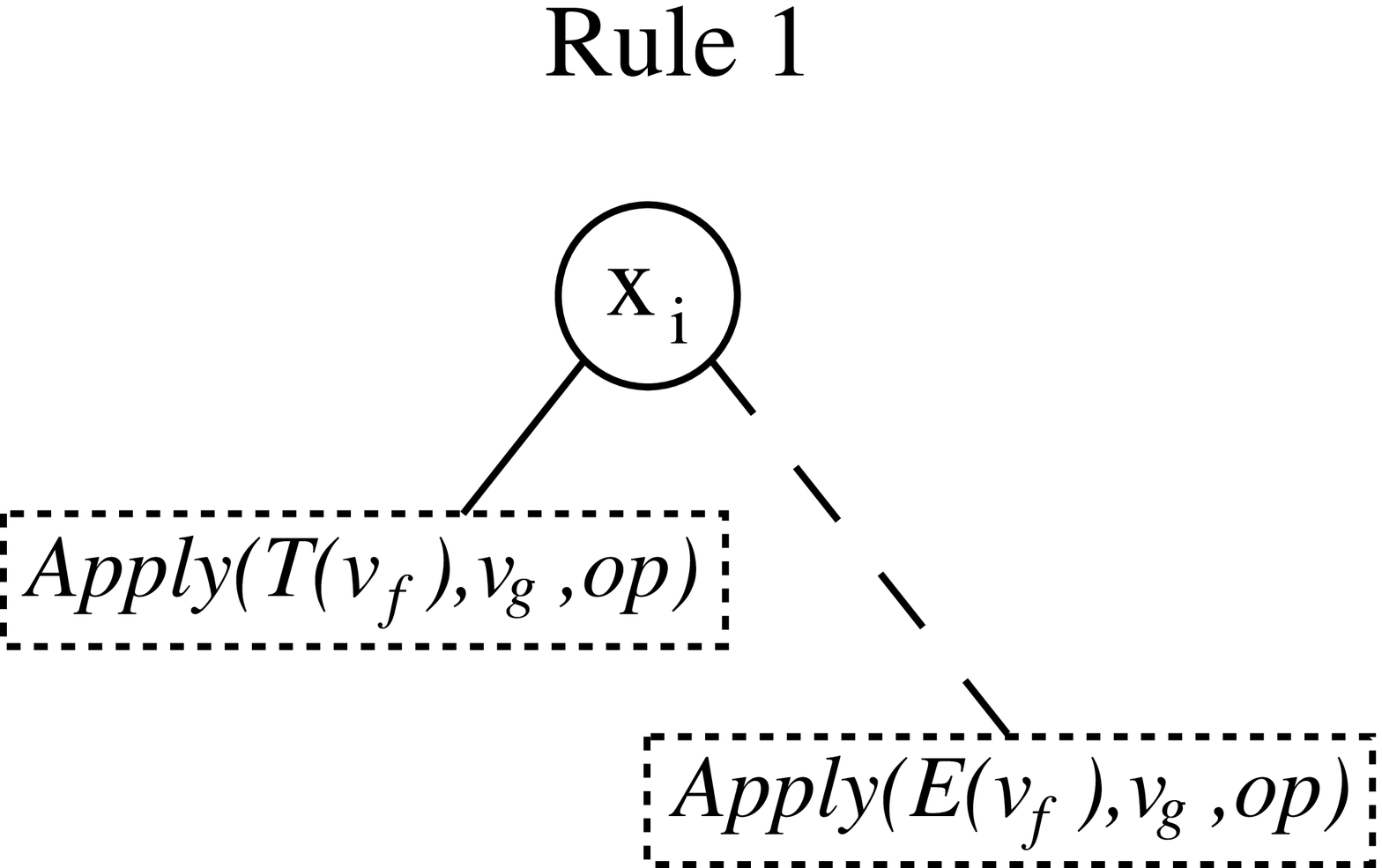}
      &
      \includegraphics[width=4.7cm]{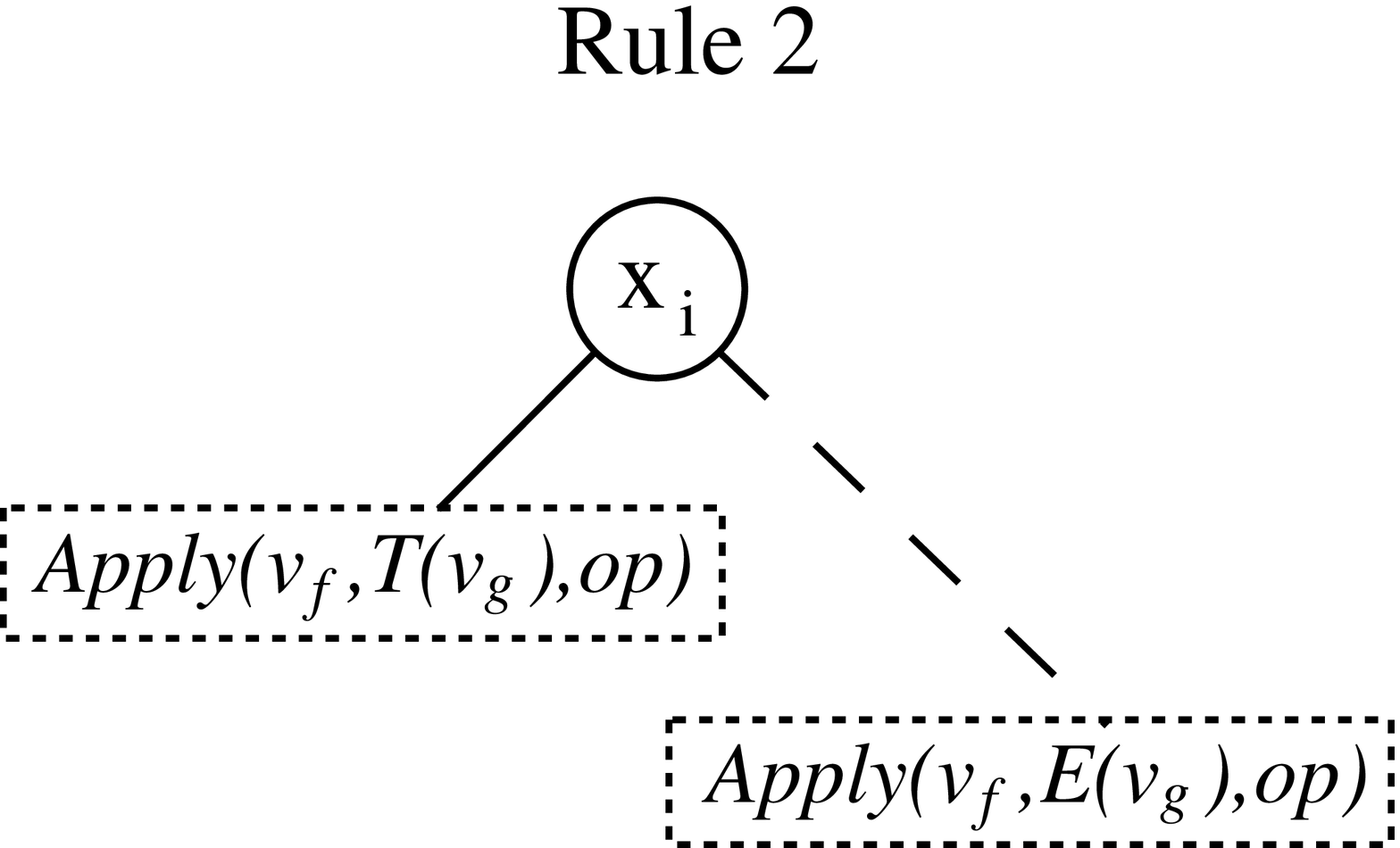}
      &
      \includegraphics[width=5.4cm]{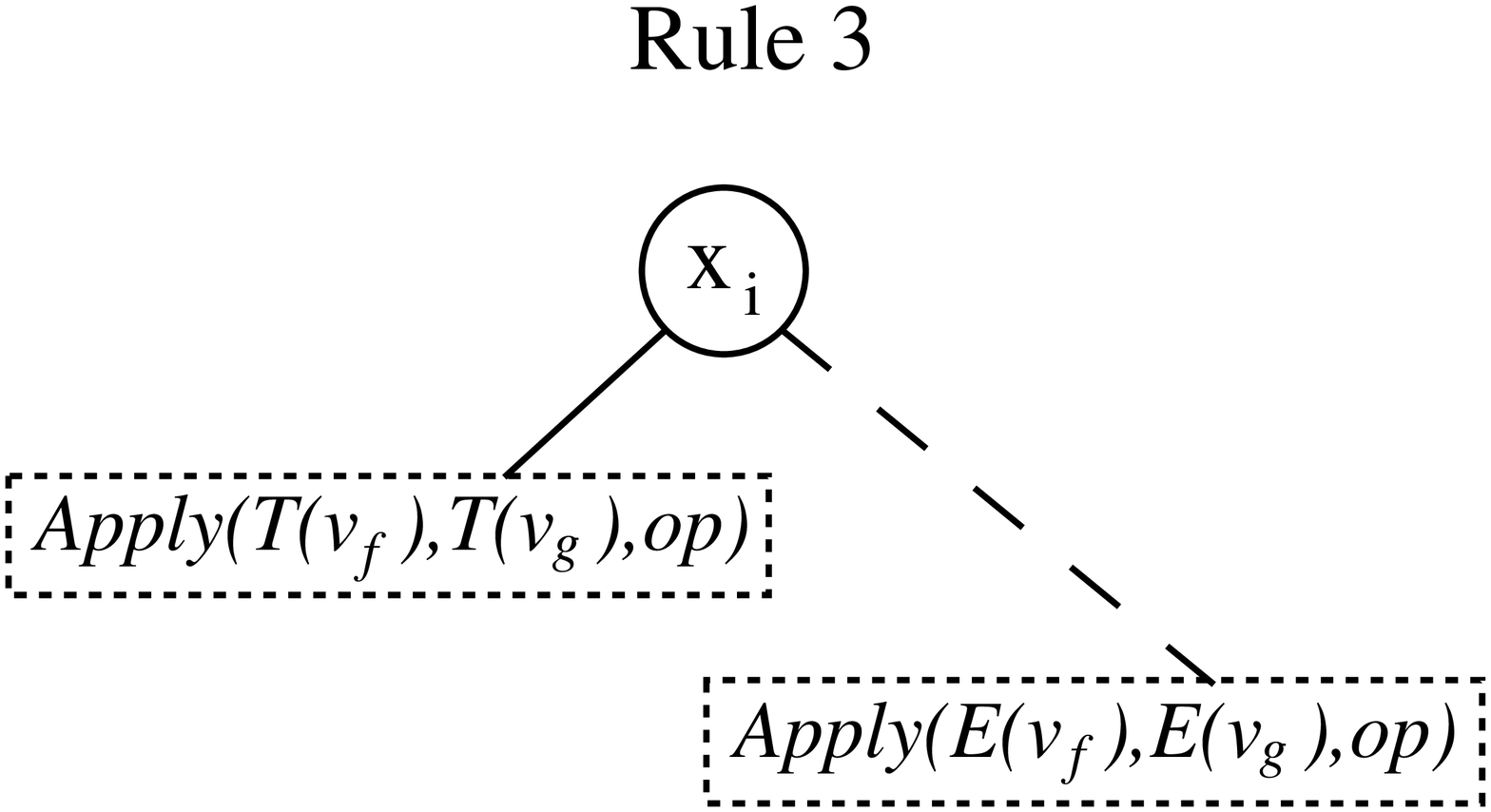}
      \\
      $x_i \prec x_j$
      &
      $x_i \succ x_j$
      &
      $x_i = x_j$
    \end{tabular}
    \parbox{14cm}{\caption{\label{fig:apply_rules} The three recursive
    rules used by the $Apply$ operation which determine how a new node
    should be added to a resultant ROBDD. In the figure, $x_i =
    Var(v_f)$ and $x_j = Var(v_g)$. The notation $x_i \prec x_j$ is
    defined to mean that $x_i$ precedes $x_j$ in the variable
    ordering.}}
  \end{center}
\end{figure}

Clearly the rules depend on the variable ordering. To illustrate,
consider performing $Apply$ using a binary operation $op$ and two
ROBDDs $f$ and $g$. $Apply$ takes as arguments two nodes, one from $f$
and one from $g$, and the operation $op$. This is denoted as
$Apply(v_f, v_g, op)$. $Apply$ compares $Var(v_f)$ and $Var(v_g)$ and
adds a new internal node to the ROBDD result using the three
rules. The rules also guide $Apply$'s traversal of the {\em then} and
{\em else} edges (this is the recursive step). For example, suppose
$Apply(v_f, v_g, op)$ is called and $Var(v_f) \prec Var(v_g)$. Rule 1
is invoked, causing an internal node containing $Var(v_f)$ to be added
to the resulting ROBDD. Rule 1 then directs the $Apply$ operation to
call itself recursively with $Apply(T(v_f), v_g, op)$ and
$Apply(E(v_f), v_g, op)$. Rules 2 and 3 dictate similar actions but
handle the cases when $Var(v_f) \succ Var(v_g)$ and $Var(v_f) =
Var(v_g)$. To recurse over both ROBDD operands correctly, the initial
call to $Apply$ must be $Apply(Root(f), Root(g), op)$ where $Root(f)$
and $Root(g)$ are the root nodes for the ROBDDs $f$ and $g$.

The recursion stops when both $v_f$ and $v_g$ are terminal nodes.
When this occurs, $op$ is performed with the values of the
terminals as operands, and the resulting value is added to the
ROBDD result as a terminal node. For example, if $v_f$ contains
the value logical $1$, $v_g$ contains the value logical $0$, and
op is defined to be $\oplus$ ($XOR$), then a new terminal with
value $1 \oplus 0 = 1$ is added to the ROBDD result. Terminal
nodes are considered {\em after} all variables are considered.
Thus, when a terminal node is compared to an internal node, either
Rule 1 or Rule 2 will be invoked depending on which ROBDD the
internal node is from.

%


ROBDD variants have been adopted in several contexts outside the
domain of logic design. Of particular relevance to this work are
Multi-Terminal Binary Decision Diagrams (MTBDDs) \cite{Clarke96}
and Algebraic Decision Diagrams (ADDs) \cite{Bahar97}. These data
structures are compressed representations of matrices and vectors
rather than logic functions, and the amount of compression
achieved is proportional to the frequency of repeated values in a
given matrix or vector. Additionally, some standard
linear-algebraic operations, such as matrix multiplication, are
defined for MTBDDs and ADDs. Since they are based on the $Apply$
operation, the efficiency of these operations is proportional to
the size in nodes of the MTBDDs or ADDs being manipulated. Further
discussion of the MTBDD and ADD representations is deferred to
Subsection \ref{sec:vec_matrix} where the general structure of the
QuIDD is described.


\subsection{Previous Linear-Algebraic Techniques}
\label{sec:matrix_ops}

Quantum-circuit simulators must support linear-algebraic operations
such as matrix multiplication, the tensor product, and the projection
operators. They typically employ array-based methods to multiply
matrices and so require exponential computational resources in the
number of qubits. Such methods are often insensitive to the actual
values stored, and even sparse-matrix storage offers little
improvement for quantum operators with no zero matrix elements, such
as Hadamard operators.

Several clever matrix methods have been developed for quantum
simulation. For example, one can simulate $k$-input quantum gates
on an $n$-qubit state vector ($k \leq n$) without explicitly
storing a $2^{n}\times 2^{n}$-matrix representation. The basic
idea is to simulate the full-fledged matrix-vector multiplication
by a series of simpler operations. To illustrate, consider
simulating a quantum circuit in which a $1$-qubit Hadamard
operator is applied to the third qubit of the state-space $|00100
\rangle$. The state vector representing this state-space has
$2^{5}$  elements. A naive way to apply the $1$-qubit Hadamard is
to construct a  $2^{5}\times 2^{5}$  matrix of the form $I \otimes
I \otimes H \otimes I \otimes I$ and then multiply this matrix by
the state vector. However, rather than compute $(I \otimes I
\otimes H \otimes I \otimes I)|00100 \rangle$, one can simply
compute $|00 \rangle \otimes H|1 \rangle \otimes |00 \rangle$,
which produces the same result using a $2\times 2$ matrix $H$. The
same technique can be applied when the state-space is in a
superposition, such as $\alpha |00100 \rangle + \beta |00000
\rangle$. In this case, to simulate the application of a $1$-qubit
Hadamard operator to the third qubit, one can compute $|00 \rangle
\otimes H(\alpha |1 \rangle + \beta |0 \rangle ) \otimes |00
\rangle$. As in the previous example, a $2\times 2$ matrix is
sufficient.

While the above method allows one to compute a state space
symbolically, in a realistic simulation environment, state vectors
may be much more complicated. Shortcuts that take advantage of the
linearity of matrix-vector multiplication are desirable. For
example, a single qubit can be manipulated in a state vector by
extracting a certain set of two-dimensional vectors. Each vector
in such a set is composed of two probability amplitudes. The
corresponding qubit states for these amplitudes differ in value at
the position of the qubit being operated on but agree in every
other qubit position. The two-dimensional vectors are then
multiplied by matrices representing single qubit gates in the
circuit being simulated. We refer to this technique as {\em
qubit-wise multiplication} because the state-space is manipulated
one qubit at a time. Obenland implemented a technique of this kind
as part of a simulator for quantum circuits \cite{Obenland1997}.
His method applies one- and two-qubit operator matrices to state
vectors of size $2^n$. Unfortunately, in the best case where $k =
1$, this only reduces the runtime and memory complexity from
$O(2^{2n})$ to $O(2^n)$, which is still exponential in the number
of qubits.


Gottesman developed a simulation method involving the {\em
Heisenberg representation} of quantum computation which tracks the
commutators of operators applied by a quantum circuit
\cite{Gottesman98}. With
 this model, the state vector need not be
represented explicitly because the operators describe how an
arbitrary state vector would be altered by the circuit. Gottesman
showed that simulation based on this model requires only
polynomial memory and runtime on a classical computer in certain
cases. However, it appears limited to the Clifford and Pauli
groups of quantum operators, which do not form a universal gate
library.


Other advanced simulation techniques including MATLAB's ``packed''
representation, apply data compression to matrices and vectors,
but cannot perform matrix-vector multiplication on compressed
matrices and vectors. A notable exception is Greve's simulation of
Shor's algorithm which uses BDDs \cite{shornuf}.
Probability amplitudes of individual qubits are modeled by single
decision nodes.  This only captures superpositions where every
participating qubit is rotated by $\pm 45$ degrees from
$|0\rangle$ toward $|1\rangle$.
Another BDD-based technique was recently proposed by Al-Rabadi et
al. \cite{Alrabadi2002} which can perform multi-valued quantum
logic. A drawback of this technique is that it is limited to 
{\em synthesis} of quantum logic gates rather than simulation
of their behavior.


Though Greve's and Al-Rabadi et al.'s BDD representations cannot
simulate  arbitrary quantum circuits, the idea of modeling quantum
states with a BDD-based structure is appealing and motivates our
approach.  Unlike previous techniques, this approach is
capable of simulating arbitrary quantum circuits while offering
performance improvements as demonstrated by the  results presented
in Sections \ref{sec:qproof} and \ref{sec:exp}.

\section{QuIDD Theory}
\label{sec:qtheory}

The {\em Quantum Information Decision Diagram} (QuIDD) was born out of
the observation that vectors and matrices which arise in quantum
computing exhibit repeated structure. Complex operators obtained from
the tensor product of simpler matrices continue to exhibit common
substructures which certain BDD variants can capture. MTBDDs and ADDs,
introduced in Subsection \ref{sec:bdd_ops}, are particularly relevant
to the task of simulating quantum systems. The QuIDD can be viewed as
an ADD or MTBDD with the following properties:

\begin{enumerate}

{\item The values of terminal nodes are restricted to the set of
complex numbers}

{\item Rather than contain the values explicitly, QuIDD terminal nodes
  contain integer indices which map into a separate array of complex
  numbers. This allows the use of a simpler integer function for
  $Apply$-based operations, along with existing ADD and MTBDD
  libraries \cite{cudd}, greatly reducing implementation overhead.}


{\item The variable ordering of QuIDDs interleaves row and column
  variables, which favors compression of block patterns (see
  Subsection \ref{sec:var_order})}

{\item Bahar et al. note that ADDs can be padded with 0's to represent
  arbitrarily sized matrices \cite{Bahar97}. No such padding is
  necessary in the quantum domain where all vectors and matrices have
  sizes that are a power of 2 (see Subsection \ref{sec:var_order})}

\end{enumerate}
As we demonstrate using our QuIDD-based simulator 
QuIDDPro these properties greatly enhance performance of quantum
computational simulation.

\begin{figure}[t]
\begin{center}
  \includegraphics[width=12cm]{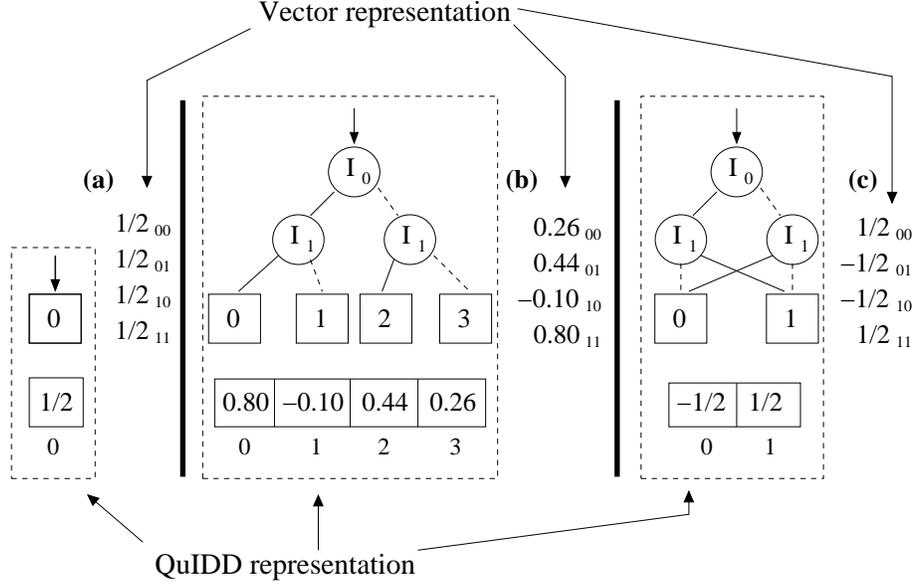}

  \parbox{12cm}{\caption{\label{fig:compress} Sample QuIDDs for state
      vectors of (a) best, (b) worst and (c) mid-range size.}}
\end{center}
\vspace{-8mm}
\end{figure}

\subsection{Vectors and Matrices}
\label{sec:vec_matrix}

Figure \ref{fig:compress} shows the QuIDD structure for three
$2$-qubit states. We consider the indices of the four vector elements
to be binary numbers, and define their bits as decision variables of
QuIDDs.  A similar definition is used for ADDs \cite{Bahar97}. For
example, traversing the {\it then} edge (solid line) of node $I_0$ in
Figure \ref{fig:compress}c is equivalent to assigning the value $1$ to
the first bit of the 2-bit vector index. Traversing the {\it else}
edge (dotted line) of node $I_1$ in the same figure is equivalent to
assigning the value $0$ to the second bit of the index. These
traversals bring us to the terminal value $-\frac{1}{2}$, which is
precisely the value at index $10$ in the vector representation.

\begin{figure*}[tb]
  \begin{center}
    \begin{tabular}{cc}
     \parbox{7cm}
     {
       {\Large
       \vspace{-44mm}
      $
      \begin{array} {cc}
    R_0R_1 & \\
        \begin{array} {r}
          _{00} \\ \\ _{01} \\ \\ _{10} \\ \\ _{11}
        \end{array} &
        \left[
          \begin{array}{rrrr}
            \frac{1}{2} & \frac{1}{2} & \frac{1}{2} & \frac{1}{2} \\ \\
            \frac{1}{2} & -\frac{1}{2} & \frac{1}{2} & -\frac{1}{2} \\ \\
            \frac{1}{2} & \frac{1}{2} & -\frac{1}{2} & -\frac{1}{2} \\ \\
            \frac{1}{2} & -\frac{1}{2} & -\frac{1}{2} & \frac{1}{2}
          \end{array}
        \right] \\ &
        \begin{array} {llll}
          _{00}~~~ & _{01}~~ & _{10}~ & ~~_{11} \\
          \multicolumn{4}{c}{C_0C_1} \\
          \multicolumn{4}{c}{\textrm{(a)}}
        \end{array}
      \end{array}
      $
      }
     }
      &
      \hspace{-3mm}\includegraphics[width=8cm]{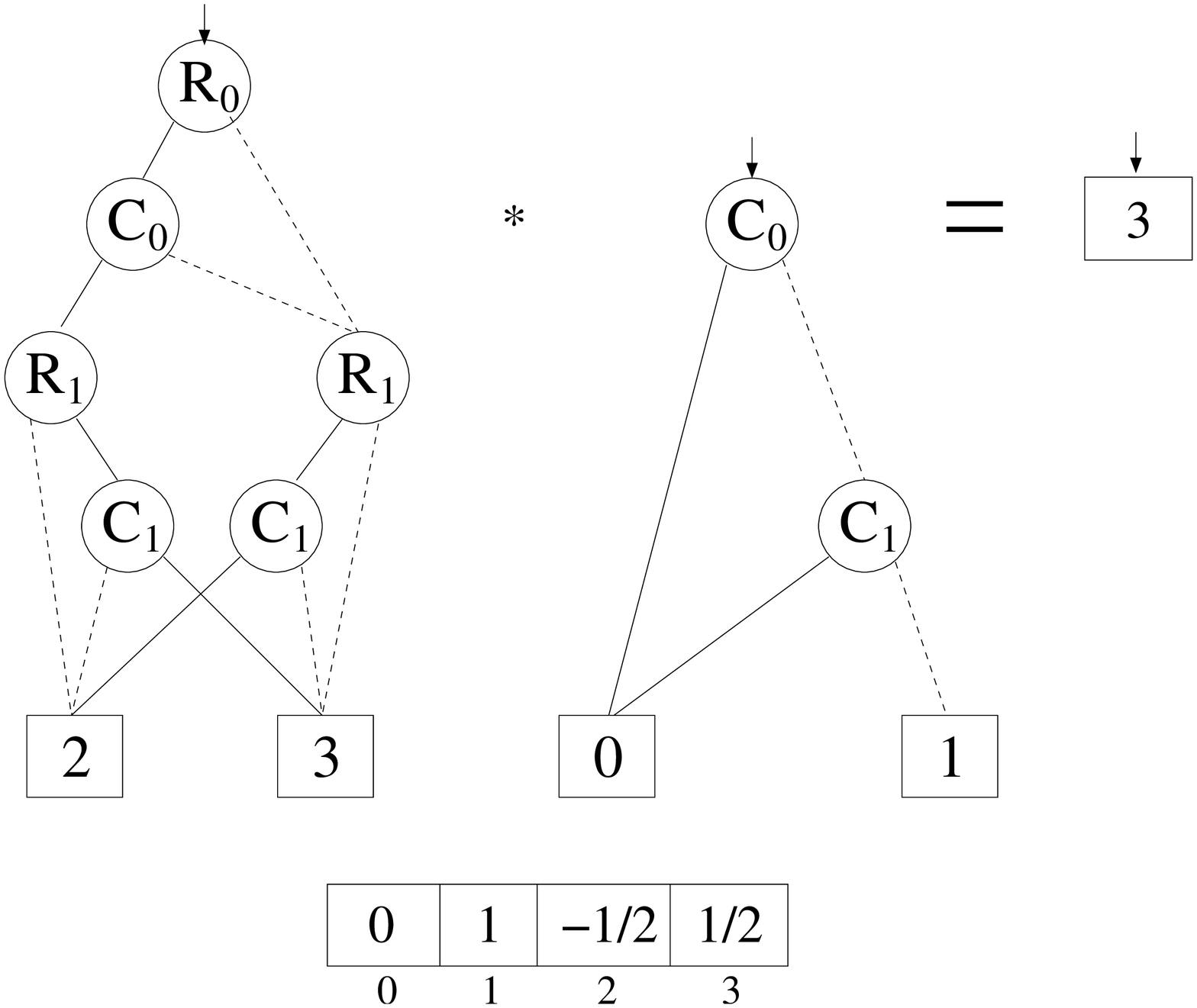}
      \vspace{-5mm} \\
       & {\Large(b)}

      \\
    \end{tabular}
    \parbox{14cm}{\caption{\label{fig:mmult} (a) $2$-qubit Hadamard,
        and (b) its QuIDD representation multiplied by $|00 \rangle =
        (1,0,0,0)$. Note that the vector and matrix QuIDDs share the
        entries in a terminal array that is global to the computation.}}
  \end{center}
  \vspace{-6mm}
\end{figure*}

QuIDD representations of matrices extend those of vectors by adding a
second type of variable node and enjoy the same reduction rules and
compression benefits. Consider the $2$-qubit Hadamard matrix annotated
with binary row and column indices shown in Figure \ref{fig:mmult}a.
In this case there are two sets of indices: The first (vertical) set
corresponds to the rows, while the second (horizontal) set corresponds
to the columns. We assign the variable name $R_i$ and $C_i$ to the row
and column index variables respectively. This distinction between the
two sets of variables was originally noted in several works including
that of Bahar et al. \cite{Bahar97}. Figure \ref{fig:mmult}b shows the
QuIDD form of this sample matrix where it is used to modify the state
vector $|00 \rangle = (1,0,0,0)$ via matrix-vector multiplication, an
operation discussed in more detail in Subsection \ref{sec:mult}.

\subsection{Variable Ordering}
\label{sec:var_order}

As explained in Subsection \ref{sec:bdd}, variable ordering can
drastically affect the level of compression achieved in BDD-based
structures such as QuIDDs. The CUDD programming library \cite{cudd},
which is incorporated into QuIDDPro, offers sophisticated dynamic
variable-reordering techniques that achieve performance improvements
in various BDD applications. However, dynamic variable reordering has
significant time overhead, whereas finding a good static ordering in
advance may be preferable in some cases. Good variable orderings are
highly dependent upon the structure of the problem at hand, and
therefore one way to seek out a good ordering is to study the problem
domain. In the case of quantum computing, we notice that all matrices
and vectors contain $2^n$ elements where $n$ is the number of qubits
represented.  Additionally, the matrices are square and non-singular
\cite{Nielsen2000}.

McGeer et al. demonstrated that ADDs representing certain rectangular
matrices can be operated on efficiently with interleaved row and
column variables \cite{McGeer93}. Interleaving implies the following
variable ordering: $R_0 \prec C_0 \prec R_1 \prec C_1 \prec ... \prec
R_n \prec C_n$. Intuitively, the interleaved ordering causes
compression to favor regularity in block sub-structures of the
matrices.
We observe that such regularity is created by tensor products that are
required to allow multiple quantum gates to operate in parallel and
also to extend smaller quantum gates to operate on larger numbers of
qubits. The tensor product $A \otimes B$ multiplies each element of
$A$ by the whole matrix $B$ to create a larger matrix which has
dimensions $M_A \cdot M_B$ by $N_A \cdot N_B$. By definition, the
tensor product will propagate block patterns in its operands. To
illustrate the notion of block patterns and how QuIDDs take advantage
of them, consider the tensor product of two one-qubit Hadamard
operators: \\

\vspace{-4mm}
\begin{center}
{\footnotesize
$\begin{array}{ccccc}
  \left[
    \begin{array}{c|c}
      (1/\sqrt{2}) & (1/\sqrt{2}) \\
      \hline
      (1/\sqrt{2}) & -1/\sqrt{2}
    \end{array}
    \right]
  &
  \hspace{-3mm}\otimes
  &
  \hspace{-3mm}\left[
    \begin{array}{c|c}
      (1/\sqrt{2}) & (1/\sqrt{2}) \\
      \hline
      (1/\sqrt{2}) & -1/\sqrt{2}
    \end{array}
    \right]
  &
  =
  &
  \left[
    \begin{array}{c|c}
      \left(
      \begin{array}{cc}
    1/2 & 1/2 \\
    1/2 & -1/2
      \end{array}
      \right)
      &
      \left(
      \begin{array}{cc}
    1/2 & 1/2 \\
    1/2 & -1/2
      \end{array}
      \right) \\

      \hline

      \left(
      \begin{array}{cc}
    1/2 & 1/2 \\
    1/2 & -1/2
      \end{array}
      \right)
      &
      \begin{array}{cc}
    -1/2 & -1/2 \\
    -1/2 & 1/2
      \end{array}
    \end{array}
    \right]
\end{array}$ \\
}
\end{center}

\noindent The above matrices have been separated into quadrants, and
each quadrant represents a block. For the Hadamard matrices depicted,
three of the four blocks are equal in both of the one-qubit matrices
and also in the larger two-qubit matrix (the equivalent blocks are
surrounded by parentheses). This repetition of equivalent blocks
demonstrates that the tensor product of two equal matrices propagates
block patterns. In the case of the above example, the pattern is that
all but the lower-right quadrant of an n-qubit Hadamard operator are
equal. Furthermore, the structure of the two-qubit matrix implies a
recursive block sub-structure, which can be seen by recursively
partitioning each of the quadrants in the two-qubit matrix: \\

\vspace{-4mm}
\begin{center}
{\footnotesize
$\begin{array}{ccccc}
  \left[
    \begin{array}{c|c}
      (1/\sqrt{2}) & (1/\sqrt{2}) \\
      \hline
      (1/\sqrt{2}) & -1/\sqrt{2}
    \end{array}
    \right]
  &
  \hspace{-3mm}\otimes
  &
  \hspace{-3mm}\left[
    \begin{array}{c|c}
      (1/\sqrt{2}) & (1/\sqrt{2}) \\
      \hline
      (1/\sqrt{2}) & -1/\sqrt{2}
    \end{array}
    \right]
  &
  \hspace{-2mm}=
  &
  \hspace{-2mm}\left[
    \begin{array}{c|c}
      \left(
      \begin{array}{c|c}
    (1/2) & (1/2) \\
    \hline
    (1/2) & -1/2
      \end{array}
      \right)
      &
      \left(
      \begin{array}{c|c}
    (1/2) & (1/2) \\
    \hline
    (1/2) & -1/2
      \end{array}
      \right) \\ \\

      \hline \\

      \left(
      \begin{array}{c|c}
    (1/2) & (1/2) \\
    \hline
    (1/2) & -1/2
      \end{array}
      \right)
      &
      \begin{array}{c|c}
    (-1/2) & (-1/2) \\
    \hline
    (-1/2) & 1/2
      \end{array}
    \end{array}
    \right]
  \end{array}$ \\
}
\end{center}

The only difference between the values in the two-qubit matrix and
the values in the one-qubit matrices is a factor of $1/\sqrt{2}$.
Thus, we
can recursively define the Hadamard operator as follows: \\

\begin{center}
$\begin{array}{ccccc}
  H^{n-1}
  &
  \otimes
  &
  H^{n-1}
  &
  =
  &
  \left[
  \begin{array}{cc}
    C_1 H^{n-1} & C_1 H^{n-1} \\
    C_1 H^{n-1} & C_2 H^{n-1}
  \end{array}
  \right]
\end{array}$ \\
\end{center}

\noindent
where $C_1 = 1/\sqrt{2}$ and $C_2 = -1/\sqrt{2}$. Other operators
constructed via the tensor product can also be defined recursively in
a similar fashion.

Since three of the four blocks in an n-qubit Hadamard operator are
equal, significant redundancy is exhibited. The interleaved variable
ordering property allows a QuIDD to explicitly represent only two
distinct blocks rather than four as shown in Figure
\ref{fig:abs_hadamard}. As we demonstrate in Sections \ref{sec:qproof}
and \ref{sec:exp}, compression of equivalent block sub-structures
using QuIDDs offers major performance improvements for many of the
operators that are frequently used in quantum computation. In the next
Subsection, we describe an algorithm which implements the tensor
product for QuIDDs and leads to the compression just described.

\begin{figure}[t]
  \begin{center}
    \includegraphics[width=10cm]{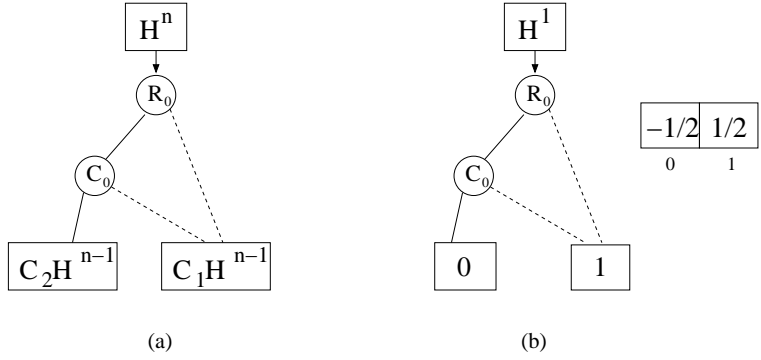}

    \parbox{12cm}{\caption{\label{fig:abs_hadamard} (a) $n$-qubit Hadamard
    QuIDD depicted next to (b) $1$-qubit Hadamard QuIDD. Notice that they
    are isomorphic except at the terminals.}}
  \end{center}
  \vspace{-5mm}
\end{figure}

\subsection{Tensor Product}
\label{sec:tensor}


With the structure and variable ordering in place, operations
involving QuIDDs can now be defined. Most operations defined for ADDs
also work on QuIDDs with some modification to accommodate the QuIDD
properties. The tensor (Kronecker) product has been described by
Clarke et al. for MTBDDs representing various arithmetic transform
matrices \cite{Clarke96}. Here we reproduce an algorithm for the
tensor product of QuIDDs based on the {\em Apply} operation that bears
similarity to Clarke's description.  Recall that the tensor product $A
\otimes B$ produces a new matrix which multiplies each element of $A$
by the entire matrix $B$.  Rows (columns) of the tensor product matrix
are component-wise products of rows (columns) of the argument
matrices. Therefore it is straightforward to implement the tensor
product operation on QuIDDs using the {\em Apply} function with an
argument that directs {\em Apply} to multiply when it reaches the
terminals of both operands. However, the main difficulty here lies in
ensuring that the terminals of $A$ are each multiplied by {\em all}
the terminals of $B$. From the definition of the standard recursive
{\em Apply} routine, we know that variables which precede other
variables in the ordering are expanded first \cite{bryant, Clarke96}.
Therefore, we must first shift all variables in $B$ in the current
order {\em after} all of the variables in $A$ prior to the call to
{\em Apply}. After this shift is performed, the {\em Apply} routine
will then produce the desired behavior. {\em Apply} starts out with $A
* B$ and expands $A$ alone until $A_{terminal} * B$ is reached for
each terminal in $A$.  Once a terminal of $A$ is reached, $B$ is fully
expanded, implying that each terminal of $A$ is multiplied by all of
$B$.
The size of the resulting QuIDD and the runtime for generating it
given two operands of sizes $a$ and $b$ (in number of nodes) is
$O(ab)$ because the tensor product simply involves a variable shift of
complexity $O(b)$, followed by a call to {\em Apply}, which Bryant
showed to have time and memory complexity $O(ab)$ \cite{bryant}.

\subsection{Matrix Multiplication}
\label{sec:mult}

Matrix multiplication can be implemented very efficiently by using
$Apply$ to implement the dot-product operation. This follows from the
observation that multiplication is a series of dot-products between
the rows of one operand and the columns of the other operand. In
particular, given matrices $A$ and $B$ with elements $a_{ij}$ and
$b_{ij}$, their product $C=AB$ can be computed element-wise by $c_{ij}
= \Sigma _{j=1}^{n} a_{ij}b_{ji}$.

Matrix multiplication for QuIDDs is an extension of the {\em Apply}
function that implements the dot-product. One call to {\em Apply} will
not suffice because the dot-product requires {\em two} binary
operations to be performed, namely addition and multiplication. To
implement this we simply use the matrix multiplication algorithm
defined by Bahar et al. for ADDs \cite{Bahar97} but modified
to support the QuIDD properties. The algorithm essentially makes two
calls to {\em Apply}, one for multiplication and the other for
addition.

Another important issue in efficient matrix multiplication is
compression. To avoid the same problem that MATLAB encounters with
its ``packed'' representation, ADDs do not require decompression
during matrix multiplication. In the work of Bahar et al., this is
addressed by tracking the number $i$ of ``skipped'' variables
between the parent node and its child node in each recursive call.
To illustrate, suppose that $Var(v_f) = x_2$ and $Var(T(v_f)) =
x_5$. In this situation, $i = 5 - 2 = 3$. A factor of $2^i$ is
multiplied by the terminal-terminal product that is reached at the
end of a recursive traversal \cite{Bahar97}.

The pseudo-code presented for this algorithm in subsequent work of
Bahar et al. suggests time-complexity $O((ab)^2)$ where $a$ and $b$
are the sizes, i.e., the number of decision nodes, of two ADD operands
\cite{Bahar97}. As with all BDD algorithms based on the {\em Apply}
function, the size of the resulting ADD is on the order of the time
complexity, meaning that the size is also $O((ab)^2)$. In the context
of QuIDDs, we use a modified form of this algorithm to multiply
operators by the state vector, meaning that $a$ and $b$ will be the
sizes in nodes of a QuIDD matrix and QuIDD state vector,
respectively. If either $a$ or $b$ or both are exponential in the
number of qubits in the circuit, the QuIDD approach will have
exponential time and memory complexity.  However, in Section
\ref{sec:qproof} we formally argue that many of the operators which
arise in quantum computing have QuIDD representations that are
polynomial in the number of qubits.

Two important modifications must be made to the ADD matrix multiply
algorithm in order to adapt it for QuIDDs. To satisfy QuIDD properties
1 and 2, the algorithm must treat the terminals as indices into an
array rather than the actual values to be multiplied and added. Also,
a variable ordering problem must be accounted for when multiplying a
matrix by a vector. A QuIDD matrix is composed of interleaved row and
column variables, whereas a QuIDD vector only depends on column
variables. If the ADD algorithm is run as described above without
modification, the resulting QuIDD vector will be composed of row
instead of column variables. The structure will be correct, but the
dependence on row variables prevents the QuIDD vector from being used
in future multiplications. Thus, we introduce a simple extension which
transposes the row variables in the new QuIDD vector to corresponding
column variables. In other words, for each $R_i$ variable that exists
in the QuIDD vector's support, we map that variable to $C_i$.



\subsection{Other Linear-Algebraic Operations}

Matrix addition is easily implemented by calling {\em Apply} with $op$
defined to be addition. Unlike the tensor product, no special variable
order shifting is required for matrix addition. Another interesting
operation which is nearly identical to matrix addition is element-wise
multiplication $c_{ij} = a_{ij}b_{ij}$. Unlike the dot-product, this
operation involves only products and no summation. This algorithm is
implemented just like matrix addition except that $op$ is defined to
be multiplication rather than addition. In quantum computer
simulation, this operation is useful for matrix-vector multiplications
with a diagonal matrix like the Conditional Phase Shift in Grover's
algorithm \cite{Grover97}. Such a shortcut considerably improves upon
full-fledged matrix multiplication.  Interestingly enough,
element-wise multiplication, and matrix addition operations for QuIDDs
can perform, without the loss of efficiency, respective scalar
operations. That is because a QuIDD with a single terminal node can be
viewed both as a scalar value and as a matrix or vector with repeated
values.

 Since matrix addition, element-wise multiplication, and their scalar
counterparts are nothing more than calls to {\em Apply}, the runtime
complexity of each operation is $O(ab)$ where $a$ and $b$ are the
sizes in nodes of the QuIDD operands. Likewise, the resulting QuIDD
has memory complexity $O(ab)$ \cite{bryant}.

Another relevant operation which can be performed on QuIDDs is the
transpose. It is perhaps the simplest QuIDD operation because it
is accomplished by swapping the row and column variables of a
QuIDD. The transpose is easily extended to the complex conjugate
transpose\footnote{The complex conjugate transpose is also known
as the Hermitian conjugate or the adjoint.} by first performing
the transpose of a QuIDD and then conjugating its terminal values.
The runtime and memory complexity of these operations is $O(a)$
where $a$ is the size in nodes of the QuIDD undergoing a
transpose.

To perform quantum measurement (see Subsection
\ref{sec:measurement}) one can use the inner product, which can be
faster than multiplying by projection matrices and computing
norms. Using the transpose, the inner product can be defined for
QuIDDs.  The inner product of two QuIDD vectors, e.g., $\langle
A|B \rangle$, is computed by matrix multiplying the transpose of
$A$ with $B$. Since matrix multiplication is involved, the runtime
and memory complexity of the inner product is $O((ab)^2)$, where
$a$ and $b$ are the sizes in nodes of $A$ and $B$ respectively.
Our current QuIDD-based simulator QuIDDPro supports matrix
multiplication, the tensor product, measurement, matrix addition,
element-wise multiplication, scalar operations, the transpose, the
complex conjugate transpose, and the inner product.

\subsection{Measurement}
\label{sec:measurement}

Measurement can be defined for QuIDDs using a combination of
operations. After measurement, the state vector is described by:

\begin{center}

  $\frac{M_m | \psi \rangle }{\sqrt{ \langle \psi | M_m^{ \dagger } M_m | \psi \rangle }}$

\end{center}

\noindent$M_m$ is a measurement operator and can be represented by a
QuIDD matrix, and the state vector $| \psi \rangle$ can be represented
by a QuIDD vector. The expression in the numerator involves a QuIDD
matrix multiplication. In the denominator, $M_m^{ \dagger }$ is the
complex conjugate transpose of $M_m$, which is also defined for
QuIDDs. $M_m^{ \dagger } M_m$ and $M_m^{ \dagger } M_m | \psi \rangle$
are matrix multiplications. $\langle \psi | M_m^{ \dagger } M_m | \psi
\rangle$ is an inner product which produces a QuIDD with a single
terminal node. Taking the square root of the value in this terminal
node is straightforward. To complete the measurement, scalar division
is performed with the QuIDD in the numerator and the single terminal
QuIDD in the denominator as operands.

Let $a$ and $b$ be the sizes in nodes of the measurement operator
QuIDD and state vector QuIDD, respectively. Performing the matrix
multiplication in the numerator has runtime and memory complexity
$O((ab)^2)$. The scalar division between the numerator and denominator
also has the same runtime and memory complexity since the denominator
is a QuIDD with a single terminal node. However, computing the
denominator will have runtime and memory complexity $O(a^{16}b^6)$ due
to the matrix-vector multiplications and inner product.

\section{Complexity Analyses}
\label{sec:qproof}

In this section we prove that the QuIDD data structure can
represent a large class of state vectors and operators using an
amount of memory that is {\em linear} in the number of qubits
rather than exponential. Further, we prove that the QuIDD
operations required in quantum circuit simulation, i.e., matrix
multiplication, the tensor product, and measurement, have both
runtime and memory that is linear in the number of qubits for the
same class of state vectors and operators. In addition to these
complexity issues, we also analyze the runtime and memory
complexity of simulating Grover's algorithm using QuIDDs.

\subsection{Complexity of QuIDDs and QuIDD Operations}

The key to analyzing the runtime and memory complexity of the
QuIDD-based simulations lies in describing the mechanics of the tensor
product. Indeed, the tensor product is the means by which quantum
circuits can be represented with matrices. In the following analysis,
the size of a QuIDD is represented by the number of nodes rather than
actual memory consumption. Since the amount of memory used by a single
QuIDD node is a constant, size in nodes is relevant for asymptotic
complexity arguments. Actual memory usage in megabytes of QuIDD
simulations is reported in Section \ref{sec:exp}.

Figure \ref{fig:theory} illustrates the general form of a tensor
product between two QuIDDs $A$ and $B$. $In(A)$ represents the
internal nodes of $A$, while $a_1$ through $a_x$ denote terminal
nodes. The notation for $B$ is similar.

\begin{figure}[ht]
  \begin{center}
    \includegraphics[width=16cm, height=10cm]{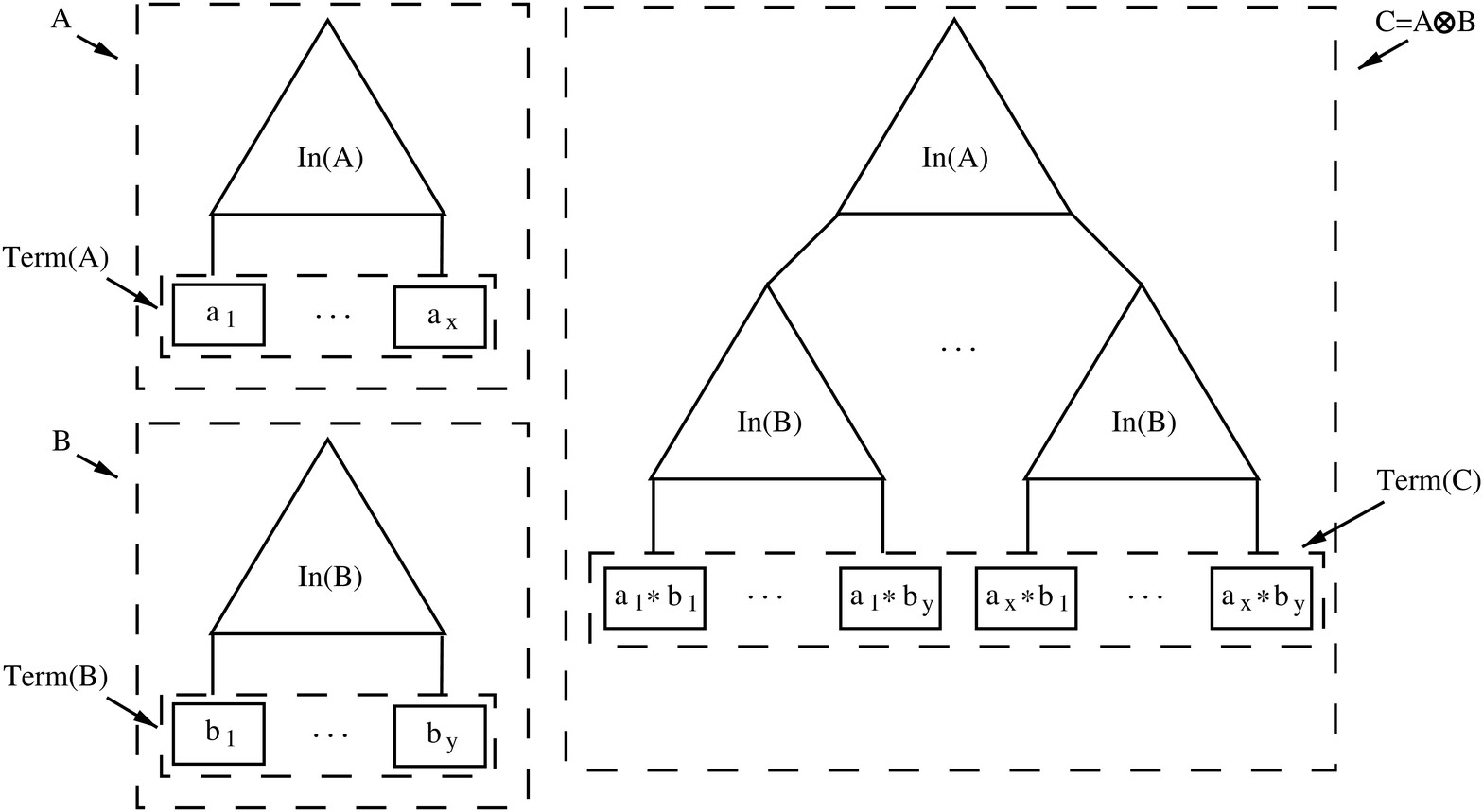}
    \parbox{12cm}
    {
      \caption{ \label{fig:theory}
      General form of a tensor product between two QuIDDs $A$ and $B$.}
    }
  \end{center}
\end{figure}

$In(A)$ is the root subgraph of the tensor product result because of
the interleaved variable ordering defined for QuIDDs and the variable
shifting operation of the tensor product (see Subsection
\ref{sec:tensor}). Suppose that $A$ depends on the variables $R_0
\prec C_0 \prec \ldots \prec R_i \prec C_i$, and $B$ depends on the
variables $R_0 \prec C_0 \prec \ldots \prec R_j \prec C_j$. In
performing $A \otimes B$, the variables on which $B$ depends will be
shifted to $R_{i+1} \prec C_{i+1} \prec \ldots \prec R_{k+i+1} \prec
C_{k+i+1}$. The tensor product is then completed by calling $Apply(A,
B, *)$. Due to the variable shift on $B$, Rule 1 of the $Apply$
function will be used after each comparison of a node from $A$ with a
node from $B$ until the terminals of $A$ are reached. Using Rule 1 for
each of these comparisons implies that only nodes from $A$ will be
added to the result, explaining the presence of $In(A)$. Once the
terminals of $A$ are reached, Rule 2 of $Apply$ will then be invoked
since terminals are defined to appear last in the variable
ordering. Using Rule 2 when the terminals of $A$ are reached implies
that all the internal nodes from $B$ will be added in place of each
terminal of $A$, causing $x$ copies of $In(B)$ to appear in the result
(recall that there are $x$ terminals in $A$). When the terminals of
$B$ are reached, they are multiplied by the appropriate terminals of
$A$. Specifically, the terminals of a copy of $B$ will each be
multiplied by the terminal of $A$ that its $In(B)$ replaced. The same
reasoning holds for QuIDD vectors as vectors differ in that they
depend only on $R_i$ variables.

Figure \ref{fig:theory} suggests that the size of a QuIDD constructed
via the tensor product depends on the number of terminals in the
operands. The more terminals a left-hand tensor operand contains, the
more copies of the right-hand tensor operand's internal nodes will be
added to the result. More formally, consider the tensor product of a
series of QuIDDs $\otimes_{i=1}^n Q_i = ( \ldots ((Q_1 \otimes Q_2)
\otimes Q_3) \otimes \ldots \otimes Q_n)$. Note that the $\otimes$
operation is associative (thus parenthesis do not affect the result),
but it is not commutative. The number of nodes in this tensor product
is described by the following lemma.

\begin{lemma}
  \label{lemma:formula}
  Given QuIDDs $\{Q_i\}_{i=1}^n$, the tensor-product QuIDD
  $\otimes_{i=1}^n Q_i$ contains \\ $|In(Q_1)| +
\Sigma _{i=2}^{n}|In(Q_i)||Term(\otimes _{j=1}^{i-1} Q_j)| +
|Term(\otimes _{i=1}^n Q_i)|$ nodes.\footnote{$|In(A)|$ denotes the
number of internal nodes in $A$, while $|Term(A)|$ denotes the
number of terminal nodes in $A$.}
\end{lemma}
{\bf Proof.} This formula can be verified by induction. For the base
case, $n=1$, there is a single QuIDD $Q_1$. Putting this information
into the formula eliminates the summation term, leaving $|In(Q_1)| +
|Term(Q_1)|$ as the total number of nodes in $Q_1$. This is clearly
correct since, by definition, a QuIDD is composed of its internal and
terminal nodes. To complete the proof, we now show that if the formula
is true for $Q_n$ then it's true for $Q_{n+1}$. The inductive
hypothesis for $Q_n$ is $|\otimes _{i=1}^n Q_i| = |In(Q_1)| + \Sigma
_{i=2}^n |In(Q_i)||Term(\otimes _{j=1}^{i-1} Q_j)| + |Term(\otimes
_{i=1}^n Q_i)|$. For $Q_{n+1}$ the number of nodes is: \\

$|(\otimes _{i=1}^n Q_i) \otimes Q_{n+1}|$ \\

$= |\otimes _{i=1}^n Q_i| - |Term(\otimes _{i=1}^n Q_i)| + |In(Q_{n+1})||Term(\otimes _{i=1}^n Q_i)| + |Term(\otimes _{i=1}^{n+1} Q_i)|$ \\

Notice that the number of terminals in $\otimes _{i=1}^n Q_i$ are
subtracted from the total number of nodes in $\otimes _{i=1}^n Q_i$
and multiplied by the number of internal nodes in $Q_{n+1}$. The
presence of these terms is due to Rule 2 of $Apply$ which dictates
that in the tensor-product $(\otimes _{i=1}^n Q_i) \otimes Q_{n+1}$,
the terminals of $\otimes _{i=1}^n Q_i$ are replaced by copies of
$Q_{n+1}$ where each copy's terminals are multiplied by a terminal
from $\otimes _{i=1}^n Q_i$. The last term simply accounts for the
total number of terminals in the tensor-product. Substituting the
inductive hypothesis made earlier for the term $|\otimes _{i=1}^n
Q_i|$ produces: \\

$|In(Q_1)| + \Sigma _{i=2}^n |In(Q_i)||Term(\otimes _{j=1}^{i-1} Q_j)|
+ |Term(\otimes _{i=1}^n Q_i)| - |Term(\otimes _{i=1}^n Q_i)| +$ \\
$|In(Q_{n+1})||Term(\otimes _{i=1}^n Q_i)| + |Term(\otimes _{i=1}^{n+1}
Q_i)|$ \\

$= |In(Q_1)| + \Sigma _{i=2}^{n+1} |In(Q_i)||Term(\otimes _{j=1}^{i-1}
Q_j)| + |Term(\otimes _{i=1}^{n+1})|$ \\

Thus the number of nodes in $Q_{n+1}$ is equal to the original formula
we set out to prove for $n+1$ and the induction is complete. $\Box$


Lemma \ref{lemma:formula} suggests that if the number of terminals
in $\otimes_{i=1} Q_i$ increases by a certain factor with each
$Q_i$, then $\otimes_{i=1}^n Q_i$ must grow exponentially in $n$.
If, however, the number of terminals stops changing, then
$\otimes_{i=1}^n Q_i$ must grow linearly in $n$. Thus, the growth
depends on matrix entries because terminals of $A \otimes B$ are
products of terminal values of $A$ by terminal values of $B$ and
repeated products are merged. If all QuIDDs $Q_i$ have terminal
values from the same set $\Gamma$, the product's terminal values
are products of elements from $\Gamma$.

\begin{definition}
\label{def:persistent}
  Consider finite non-empty sets of complex numbers $\Gamma_1$ and
  $\Gamma_2$, and define their {\em all-pairs product} as
  $\{ xy \ | \ x\in \Gamma_1,\ y\in \Gamma_2 \}$. One can
  verify that this operation is associative, and therefore the set
  $\Gamma^n$ of {\em all n-element products} is well defined for
  $n>0$. We then call a finite non-empty set $\Gamma\subset\mathbb{C}$
  {\em persistent} iff the size of $\Gamma^n$ is constant for all
  $n>0$.
\end{definition}

   For example, the set $\Gamma=\{c,-c\}$ is persistent for any $c$
   because $\Gamma^n=\{c^n,-c^n\}$.
   In general any set closed under multiplication is persistent,
   but that is not a necessary condition. In particular, for
   $c\neq 0$, the persistence of $\Gamma$ is equivalent to the
   persistence of $c\Gamma$. Another observation is that $\Gamma$
   is persistent if and only if $\Gamma\cup\{0\}$ is persistent.
   An important example of a persistent set is the set consisting of 0
   and all $n$-th degree roots of unity $\mathbb{U}_n=\{e^{2 \pi ik/n}| k=0..n-1\}$,
   for some $n$. Since roots of unity form a group, they are closed under
   multiplication and form a persistent set. In the Appendix,
   we show that every persistent set is either $c\mathbb{U}_n$
   for some $n$ and $c\neq 0$, or $\{0\}\cup c\mathbb{U}_n$.

   The importance of persistent sets is underlined by the following theorem.

\begin{theorem}
  \label{theorem:full_unity}
  Given a persistent set $\Gamma$ and a constant $C$,
  consider $n$ QuIDDs with at most $C$ nodes each and terminal values from
  a persistent set $\Gamma$. The tensor product of those QuIDDs has $O(n)$ nodes
  and can be computed in $O(n)$ time.
\end{theorem}
{\bf Proof.} The first and the last terms of the formula in Lemma
  \ref{lemma:formula} are bounded by $C$ and $|\Gamma|$ respectively.
  As the sizes of terminal sets in the middle term are bounded by
  $|\Gamma|$, the middle term is bounded by $|\Gamma|\sum_{i=2}^n
  |In(Q_i)| < |\Gamma|c$ since each $|In(Q_i)|$ is a constant.  The
  tensor product operation $A \otimes B$ for QuIDDs involves a shift
  of variables on $B$ followed by $Apply(A, B, *)$.  If $B$ is a QuIDD
  representing $n$ qubits, then $B$ depends on $O(n)$
  variables.\footnote{More accurately, $B$ depends on exactly $2n$
  variables if it is a matrix QuIDD and $n$ variables if it is a
  vector QuIDD.} This implies that the runtime of the variable shift
  is $O(n)$. Bryant proved that the asymptotic runtime and memory
  complexity of $Apply(A, B, binary\_op)$ is $O(|A||B|)$
  \cite{bryant}. Lemma \ref{lemma:formula} and the fact that we are
  considering QuIDDs with at most $C$ nodes and terminals from a
  persistent set $\Gamma$ imply that $|A| = O(n)$ and $|B| =
  O(1)$. Thus, $Apply(A, B, *)$ has asymptotic runtime and memory
  complexity $O(n)$, leading to an overall asymptotic runtime and
  memory complexity of $O(n)$ for computing $\otimes_{i=1}^n Q_i$.
  \hfill $\Box$ \\

  Importantly, the terminal values do not need to form a persistent
  set themselves for the theorem to hold. If they are {\em contained}
  in a persistent set, then the sets of all possible $m$-element
  products (i.e. $m \leq n$ for all $n$-element products in a set
  $\Gamma$) eventually stabilize in the sense that their sizes do not
  exceed that of $\Gamma$. However, this is only true for a fixed $m$
  rather than for the sets of products of $m$ elements and fewer. \\

  For QuIDDs $A$ and $B$, the matrix-matrix and matrix-vector product
  computations are not as sensitive to terminal values, but depend
  on sizes of the QuIDDs. Indeed, the memory and time complexity of
  this operation is $O(|A|^2|B|^2)$ \cite{Bahar97}.


\begin{theorem}
  Consider measuring an $n$-qubit QuIDD state vector $| \psi \rangle$
  using a QuIDD measurement operator $M$, where both $| \psi \rangle$
  and $M$ are constructed via the tensor product of an arbitrary
  sequence of $O(1)$-sized QuIDD vectors and matrices,
  respectively. If the terminal node values of the $O(1)-sized$ QuIDD
  vectors or operators are in a persistent set $\Gamma$, then the
  runtime and memory complexity of measuring the QuIDD state vector is
  $O(n^{22})$.
\end{theorem}
{\bf Proof.} In Subsection \ref{sec:measurement}, we showed that
  runtime and memory complexity for measuring a state vector QuIDD
  is $O(a^{16}b^6)$, where $a$ and $b$ be the sizes in nodes of the
  measurement operator QuIDD and state vector QuIDD,
  respectively. From Theorem \ref{theorem:full_unity}, the asymptotic
  memory complexity of both $a$ and $b$ is $O(n)$, leading to an overall
  runtime and memory complexity of $O(n^{22})$. \hfill $\Box$ \\

The class of QuIDDs described by Theorem \ref{theorem:full_unity} and
its corollaries, with terminals taken from the set $\{0\} \cup
c\mathbb{U}$, encompasses a large number of practical quantum state
vectors and operators. These include, but are not limited to, any
equal superposition of $n$ qubits, any sequence of $n$ qubits in the
computational basis states, $n$-qubit Pauli matrices, and $n$-qubit
Hadamard matrices.  The above results suggest a polynomial-sized QuIDD
representation of any quantum circuit on $n$ qubits in terms of such
gates if the number of gates is limited by a constant. In other words,
the above sufficient conditions apply if the depth (length) of the
circuit is limited by a constant.  Our simulation technique may use
polynomial memory and runtime in other circumstances as well, as shown
in the next Subsection.


\subsection{Complexity of Grover's Algorithm using QuIDDs}

\begin{figure}
\begin{center}
\includegraphics[width=13cm]{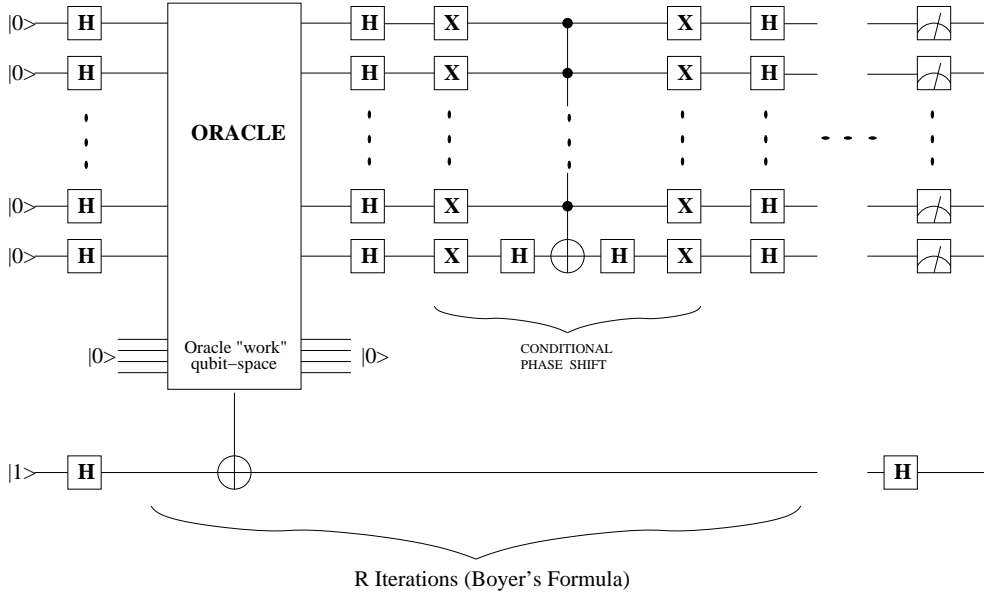}
\parbox{12cm}{\caption{\label{fig:groverckt} Circuit-level implementation of Grover's algorithm}}
\end{center}
\end{figure}

To investigate the power of the QuIDD representation, we
used QuIDDPro to simulate Grover's algorithm \cite{Grover97}, one
of the two major quantum algorithms that have been developed to
date. Grover's algorithm searches for a subset of items in an
unordered database of $N$ items. The only selection criterion
available is a black-box predicate that can be evaluated on any
item in the database. The complexity of this evaluation (query) is
unknown, and the overall complexity analysis is performed in terms
of queries. In the classical domain, any algorithm for such an
unordered search must query the predicate $\Omega(N)$ times.
However, Grover's algorithm can perform the search with quantum
query complexity $O(\sqrt{N})$, a quadratic improvement. This
assumes that a quantum version of the search predicate can be
evaluated on a superposition of all database items.

A quantum circuit representation of the algorithm involves five major
components: an \textit{oracle circuit}, a \textit{conditional phase
shift operator}, sets of Hadamard gates, the data qubits, and an
oracle qubit. The oracle circuit is a Boolean predicate that acts as a
filter, flipping the oracle qubit when it receives as input an $n$ bit
sequence representing the items being searched for. In quantum circuit
form, the oracle circuit is represented as a series of controlled NOT
gates with subsets of the data qubits acting as the control qubits and
the oracle qubit receiving the action of the NOT gates. Following the
oracle circuit, Hadamard gates put the $n$ data qubits into an equal
superposition of all $2^n$ items in the database where $2^n = N$. Then
a sequence of gates $H^{\otimes n-1} C H^{\otimes n-1}$, where $C$
denotes the conditional phase shift operator, are applied iteratively
to the data qubits. Each iteration is termed a {\em Grover iteration}
\cite{Nielsen2000}.

Grover's algorithm must be stopped after a particular number of
iterations when the probability amplitudes of the states
representing the items sought are sufficiently boosted. There must
be enough iterations to ensure a successful measurement, but after
a certain point the probability of successful measurement starts
fading, and later changes periodically. In our experiments, we
used the tight bound on the number of iterations formulated by
Boyer et al. \cite{Boyer96} when the number of solutions $M$ is
known in advance: $\lfloor \pi / 4 \theta \rfloor$ where $\theta =
\sqrt{M/N}$. The power of Grover's algorithm lies in the fact that
the data qubits store all $N = 2^n$ items in the database as a
superposition, allowing the oracle circuit to ``find'' all items
being searched for \textit{simultaneously}. A circuit implementing
Grover's algorithm is shown in Fig.  \ref{fig:groverckt}.  The
algorithm can be summarized as follows: \\

\noindent
Let $N$ denote the number of elements in the database. \\


\noindent
  {\bf 1.} Initialize $n = \lceil\log_2 N \rceil$ qubits to $|0\rangle$
  and the {\it oracle qubit} to $|1\rangle$.

\noindent
  {\bf 2.} Apply the Hadamard transform (H gate) to all qubits to put them
  into a uniform superposition of basis states.

\noindent
  {\bf 3.} \label{algo:step1} Apply the oracle circuit. The oracle
  circuit can be implemented as a series of one or more CNOT gates
  representing the search criteria. The inputs to the oracle circuit
  feed into the control portions of the CNOT gates, while the oracle
  qubit is the target qubit for all of the CNOT gates. In this way,
  if the input to this circuit satisfies the search criteria, the state
  of the oracle qubit is flipped. For a superposition of inputs, those
  input basis states that satisfy the search criteria flip the oracle qubit
  in the composite state-space. The oracle circuit uses ancillary qubits
  as its workspace, reversibly returning them to their original states
  (shown as $|0\rangle$ in Fig \ref{fig:groverckt}). These ancillary qubits
  will not be operated on by any other step in the algorithm.

\noindent
  {\bf 4.} Apply the H gate to all qubits except the oracle qubit.

\noindent
  {\bf 5.}
  Apply the {\it Conditional Phase-Shift} gate on all qubits
  except the oracle qubit. This gate negates the probability amplitude
  of the $|000\dots 0\rangle$ basis state, leaving that of the others
  unaffected. It can be realized using a combination of X, H and
  $\textrm{C}^{n-1}$-NOT gates as shown. A decomposition of the
  $\textrm{C}^{n-1}$-NOT into elementary gates is given in \cite{barenco}.

\noindent
  {\bf 6.}
   \label{algo:stepn} Apply the H gate to all gates except the oracle qubit.

\noindent
  {\bf 7.} Repeat steps
   3-6 (a single Grover iteration) $R$ times, where $ R = \lfloor\frac{\pi}{4}\sqrt{\frac{N}{M}}\rfloor$ and $M$
is the number of keys matching the search criteria \cite{Boyer96}.

\noindent
  {\bf 8.} Apply the H gate to the oracle qubit in the last iteration.
  Measure the first $n$ qubits to obtain the index of the matching key
  with high probability. \\

Using explicit vectors and matrices to simulate the above procedure
would incur memory and runtime complexities of $\Omega (2^n)$.
However, this is not necessarily the case when using QuIDDs.
To show this, we present a step-by-step complexity analysis
for a QuIDD-based simulation of the procedure.


{\bf Steps 1 and 2}
Theorem \ref{theorem:full_unity} implies that the memory and runtime
complexity of step $1$ is $O(n)$ because the initial state vector only
contains elements in $c\mathbb{U}_k \cup \{0\}$ and is constructed via
the tensor product.  Step $2$ is simply a matrix multiplication of an
$n$-qubit Hadamard matrix with the state vector constructed in step
$1$.  The Hadamard matrix has memory complexity $O(n)$ by Theorem
\ref{theorem:full_unity}. Since the state vector also has memory
complexity $O(n)$, further matrix-vector multiplication
in step $2$ has $O(n^4)$ memory and runtime complexity
because computing the product of two QuIDDs $A$ and $B$
takes $O((|A||B|)^2)$ time and memory \cite{Bahar97}. This upper-bound can be
trivially tightened, however. The function of these steps is to put the qubits
into an equal superposition. For the $n$ data qubits, this produces a QuIDD
with $O(1)$ nodes because an $n$-qubit state vector representing an equal
superposition has only one distinct element, namely $\frac{1}{2^{n/2}}$. Also,
applying a Hadamard matrix to the single oracle qubit results in a QuIDD with
$O(1)$ nodes because in the worst-case, the size of a $1$-qubit QuIDD is
clearly a contant.  Since the tensor product is based on the $Apply$ algorithm
(see Subsection \ref{sec:tensor}), the result of tensoring the QuIDD
representing the data qubits in an equal superposition with the QuIDD for the
oracle qubit is a QuIDD containing $O(1)$ nodes.

{\bf Steps 3-6} In step $3$, the state vector is matrix-multiplied
by the oracle matrix. Again, the complexity of multiplying two
arbitrary QuIDDs $A$ and $B$ is $O((|A||B|)^2)$ \cite{Bahar97}.
The size of the state vector in step $3$ is $O(1)$. If the size
of the oracle is represented by $|A|$, then the memory and runtime
complexity of step $3$ is $O(|A|^2)$.  Similarly, steps $4$,
$5$, and $6$ will have polynomial memory and runtime complexity in
terms of $|A|$ and $n$.\footnote{As noted in step $5$, the
Conditional Phase-Shift operator can be decomposed into the tensor
product of single qubit matrices, giving it memory complexity
$O(n)$.}  Thus we arrive at the $O(|A|^{16}n^{14})$ worst-case
upper-bound for the memory and runtime complexity of the
simulation at step $6$. Judging from our empirical data, this
bound is typically very loose and pessimistic.

\begin{lemma}
  \label{lemma:grov_iter}
  The memory and runtime complexity of a single Grover iteration in a
  QuIDD-based simulation is $O(|A|^{16}n^{14})$.
\end{lemma}
{\bf Proof.} Steps $3-6$ make up a single Grover iteration. Since
the memory and runtime complexity of a QuIDD-based simulation
after completing Step $6$ is $O(|A|^{16}n^{14})$, the memory and
runtime complexity of a single Grover iteration is
$O(|A|^{16}n^{14})$. $\Box$
\\

{\bf Step 7}
Step $7$ does not involve a quantum operator, but rather it repeats a
Grover iteration $R = \lfloor\frac{\pi}{4}\sqrt{\frac{N}{M}}\rfloor$
times. As a result, Step $7$ induces an exponential runtime for the
simulation, since the number of Grover iterations is a function of
$N=2^n$. This is acceptable though because an actual quantum computer
would also require exponentially many Grover iterations in order to
measure one of the matching keys with a high probability
\cite{Boyer96}. Ultimately this is the reason why Grover's algorithm
only offers a {\em quadratic} and not an exponential speedup over
classical search. Since Lemma \ref{lemma:grov_iter} shows that the
memory and runtime complexity of a single Grover iteration is
polynomial in the size of the oracle QuIDD, one might guess that the
memory complexity of Step $7$ is exponential like the
runtime. However, it turns out that the size of the state vector
does not change from iteration to iteration, as shown below.

\begin{lemma}
  \label{lemma:grov_internal}
  The internal nodes of the state vector QuIDD at the end of any
  Grover iteration $i$ are equal to the internal nodes of the state
  vector QuIDD at the end of Grover iteration $i+1$.
\end{lemma}
{\bf Proof.} Each Grover iteration increases the probability of
the states representing matching keys while simultaneously
decreasing the probability of the states representing non-matching
keys. Therefore, at the end of the first iteration, the state
vector QuIDD will have a single terminal node for all the states
representing matching keys and one other terminal node, with a
lower value, for the states representing non-matching keys (there
may be two such terminal nodes for non-matching keys, depending on
machine precision). The internal nodes of the state vector QuIDD
cannot be different at the end of subsequent Grover iterations
because a Grover iteration does not change the pattern of
probability amplitudes, but only their values.
In other words, the same matching states always point to
a terminal node whose value becomes closer to $1$ after each iteration,
while the same non-matching states always point to a terminal node (or nodes)
whose value (or values) becomes closer to $0$. $\Box$

\begin{lemma}
  \label{lemma:grov_struct}
  The total number of nodes in the state vector QuIDD at the end of
  any Grover iteration $i$ is equal to the total number of nodes in
  the state vector QuIDD at the end of Grover iteration $i+1$.
\end{lemma}
{\bf Proof.} In proving Lemma \ref{lemma:grov_internal}, we showed
that the only change in the state vector QuIDD from iteration to
iteration is the values in the terminal nodes (not the number of
terminal nodes). Therefore, the number of nodes in the state
vector QuIDD is always the same at the end of every Grover
iteration. $\Box$

\begin{corollary}
  In a QuIDD-based simulation, the runtime and memory complexity of
  any Grover iteration $i$ is equal to the runtime and memory
  complexity of Grover iteration $i+1$.
\end{corollary}

{\bf Proof.} Each Grover iteration is a series of matrix
multiplications between the state vector QuIDD and several
operator QuIDDs (Steps $3-6$). The work of Bahar et al. shows that
matrix multiplication with ADDs has runtime and memory complexity
that is determined solely by the number of nodes in the operands
(see Section \ref{sec:mult}) \cite{Bahar97}. Since the total
number of nodes in the state vector QuIDD is always the same at
the end of every Grover iteration, the runtime and memory
complexity of every Grover iteration
is the same. $\Box$ \\

Lemmas \ref{lemma:grov_internal} and \ref{lemma:grov_struct} imply
that Step $7$ does not necessarily induce memory complexity that is
exponential in the number of qubits. This important fact is captured
in the following theorem.

\begin{theorem}
  \label{theorem:grov_mem}
  The memory complexity of simulating Grover's algorithm using QuIDDs
  is polynomial in the size of the oracle QuIDD and the number of
  qubits.\footnote{We do not account for the resources required
          to construct the QuIDD of the oracle.}
\end{theorem}
{\bf Proof.} The runtime and memory complexity of a single Grover
iteration is $O(|A|^{16}n^{14})$ (Lemma \ref{lemma:grov_iter}),
which includes the initialization costs of Steps $1$ and $2$.
Also, the structure of the state vector QuIDD does not change from
one Grover iteration to the next (Lemmas \ref{lemma:grov_internal}
and \ref{lemma:grov_struct}). Thus, the overall memory complexity
of simulating Grover's algorithm with QuIDDs is
$O(|A|^{16}n^{14})$, where $|A|$ is the number of nodes in the
oracle QuIDD and $n$ is the number of qubits. $\Box$ \\

While any polynomial-time quantum computation can be simulated in
polynomial space, the commonly-used linear-algebraic simulation
requires $\Omega(2^n)$ space. Also note that the case of an oracle
searching for a unique solution (originally considered by Grover)
implies that $|A|=n$. Here, most of the searching will be done
while constructing the QuIDD of the oracle, which is an entirely
classical operation.

  As we demonstrate experimentally in Section \ref{sec:exp},
for some oracles, simulating Grover's algorithm with QuIDDs has
memory complexity $\Theta (n)$. Furthermore, simulation using
QuIDDs has worst-case runtime complexity $O(R|A|^{16}n^{14})$,
where $R$ is the number of Grover iterations as defined earlier.
If $|A|$ grows polynomially with $n$, this runtime complexity is
the same as that of an ideal quantum computer, up to a polynomial
factor.

\section{Empirical Validation}
\label{sec:exp}

This section discusses  problems that arise when
implementing a QuIDD-based simulator.  It also presents
experimental results obtained from actual simulation.

\subsection{Implementation Issues}

Full support of QuIDDs requires the use of complex
arithmetic, which can lead to serious problems if numerical
precision is not adequately addressed.

{\bf Complex Number Arithmetic.} At an abstract level, ADDs can
support terminals of any numerical type, but CUDD's implementation
of ADDs does not.
For efficiency reasons, CUDD stores node information in C {\em
union}s which are interpreted numerically for terminals and as
child pointers for internal nodes.
However, it is well-known that unions are incompatible with the
use of C++ classes because their multiple interpretations hinder
the binding of correct destructors. In particular, complex numbers
in C++ are implemented as a templated class and are incompatible
with CUDD.
This was one of the motivations for storing terminal values in an
external array (QuIDD property 2).

{\bf Numerical Precision.} Another important issue is the
precision of complex numeric types.
Over the course of repeated multiplications, the values of some
terminals may become very small and induce round-off errors if the
standard IEEE double precision floating-point types are used. This
effect worsens for larger circuits. Unfortunately, such round-off
errors can significantly affect the structure of a QuIDD by merging
terminals that are only slightly different or not merging terminals
whose values should be equal, but differ by a small computational
error. 
The use of approximate comparisons with an epsilon works in certain
cases but does not scale well, particularly for creating an equal
superposition of states (a standard operation in quantum circuits). In
an equal superposition, a circuit with $n$ qubits will contain the
terminal value $\frac{1}{2^{n/2}}$ in the state vector. With the IEEE
double precision floating-point type, this value will be rounded to
$0$ at $n=2048$, preventing the use of epsilons for approximate
comparison past $n=2048$. Furthermore, a static value for epsilon will
not work well for different sized circuits. For example, $\varepsilon
= 10^{-6}$ may work well for $n=35$, but not for $n=40$ because at
$n=40$, all values may be smaller than $10^{-6}$. Therefore, to
address the problem of precision, QuIDDPro uses an arbitrary precision
floating-point type from the GMP library \cite{GMP} with the C++
complex template. Precision is then limited to the available amount of
memory in the system.

\subsection{Results for Simulating Grover's Algorithm}

\begin{table}[tb]
\small
  \begin{center}
       \begin{tabular}{|c|c|c|c|c|c|} \hline
        Circuit &\multicolumn{2}{|c|}{Hadamards}&Conditional&\multicolumn{2}{|c|}{Oracles}\\
        Size $n$&Initial&Repeated&Phase Shift&        1 &          2  \\ \hline
        20 &     80 &     83 &        21 &      99  &        108  \\ \hline
        30 &    120 &    123 &        31 &      149 &        168  \\ \hline
        40 &    160 &    163 &        41 &      199 &        228  \\ \hline
        50 &    200 &    203 &        51 &      249 &        288  \\ \hline
        60 &    240 &    243 &        61 &      299 &        348  \\ \hline
        70 &    280 &    283 &        71 &      349 &        408  \\ \hline
        80 &    320 &    323 &        81 &      399 &        468  \\ \hline
        90 &    360 &    363 &        91 &      449 &        528  \\ \hline
       100 &    400 &    403 &       101 &      499 &        588  \\ \hline
        \end{tabular}
  \parbox{10cm}{
        \caption{\label{tab:growth}
           Size of QuIDDs (\# of nodes) for Grover's algorithm.
                  }
    }
\end{center}
\vspace{-4mm}
\end{table}

\begin{table}[!htb]
 \small
  \begin{center}
    \begin{tabular}{cc}
      \begin{tabular}{|r|c|c|c|c|} \hline
        \multicolumn{5}{|c|}{\bfseries Oracle 1: Runtime (s)}\\
        \hline\hline
        $n$ & Oct & MAT & B++ & QP \\ \hline
        10     & 80.6  & 6.64  & 0.15    & 0.33 \\ \hline
        11     & 2.65e2 & 22.5  & 0.48   & 0.54 \\ \hline
        12     & 8.36e2 & 74.2 & 1.49    & 0.83 \\ \hline
        13     & 2.75e3 & 2.55e2 & 4.70  & 1.30 \\ \hline
        14     & 1.03e4 & 1.06e3 & 14.6  & 2.01 \\ \hline
        15     & 4.82e4 & 6.76e3 & 44.7  & 3.09 \\ \hline
        16     & $>24$hrs      & $>24$hrs      & 1.35e2  & 4.79 \\ \hline
        17     & $>24$hrs      & $>24$hrs      & 4.09e2  & 7.36 \\ \hline
        18     & $>24$hrs      & $>24$hrs      & 1.23e3 & 11.3 \\ \hline
        19     & $>24$hrs      & $>24$hrs      & 3.67e3 & 17.1 \\ \hline
        20     & $>24$hrs      & $>24$hrs      & 1.09e4 & 26.2 \\ \hline
    21     & $>24$hrs      & $>24$hrs      & 3.26e4 & 39.7 \\ \hline
    22     & $>24$hrs      & $>24$hrs      & $>24$hrs & 60.5 \\ \hline
    23     & $>24$hrs      & $>24$hrs      & $>24$hrs & 92.7 \\ \hline
    24     & $>24$hrs      & $>24$hrs      & $>24$hrs & 1.40e2 \\ \hline
    25     & $>24$hrs      & $>24$hrs      & $>24$hrs & 2.08e2 \\ \hline
    26     & $>24$hrs      & $>24$hrs      & $>24$hrs & 3.12e2 \\ \hline
    27     & $>24$hrs      & $>24$hrs      & $>24$hrs & 4.72e2 \\ \hline
    28     & $>24$hrs      & $>24$hrs      & $>24$hrs & 7.07e2 \\ \hline
    29     & $>24$hrs      & $>24$hrs      & $>24$hrs & 1.08e3 \\ \hline
    30     & $>24$hrs      & $>24$hrs      & $>24$hrs & 1.57e3 \\ \hline
    31     & $>24$hrs      & $>24$hrs      & $>24$hrs & 2.35e3 \\ \hline
    32     & $>24$hrs      & $>24$hrs      & $>24$hrs & 3.53e3 \\ \hline
    33     & $>24$hrs      & $>24$hrs      & $>24$hrs & 5.23e3 \\ \hline
    34     & $>24$hrs      & $>24$hrs      & $>24$hrs & 7.90e3 \\ \hline
    35     & $>24$hrs      & $>24$hrs      & $>24$hrs & 1.15e4 \\ \hline
    36     & $>24$hrs      & $>24$hrs      & $>24$hrs & 1.71e4 \\ \hline
    37     & $>24$hrs      & $>24$hrs      & $>24$hrs & 2.57e4 \\ \hline
    38     & $>24$hrs      & $>24$hrs      & $>24$hrs & 3.82e4 \\ \hline
    39     & $>24$hrs      & $>24$hrs      & $>24$hrs & 5.64e4 \\ \hline
    40     & $>24$hrs      & $>24$hrs      & $>24$hrs & 8.23e4 \\ \hline

      \end{tabular}
      &
      \begin{tabular}{|r|c|c|c|c|} \hline
        \multicolumn{5}{|c|}{\bfseries Oracle 1: Peak Memory Usage (MB)}\\
        \hline\hline
        $n$ & Oct & MAT & B++ & QP  \\ \hline
        10 & 2.64e-2 & 1.05e-2 & 3.52e-2 & 9.38e-2 \\ \hline
        11 & 5.47e-2 & 2.07e-2 & 8.20e-2 & 0.121 \\ \hline
        12 & 0.105 & 4.12e-2 & 0.176 & 0.137 \\ \hline
        13 & 0.213 & 8.22e-2 & 0.309 & 0.137 \\ \hline
        14 & 0.426 & 0.164 & 0.559 & 0.137 \\ \hline
        15 & 0.837 & 0.328 & 1.06 & 0.137  \\ \hline
        16 & 1.74 & 0.656 & 2.06 & 0.145  \\ \hline
        17 & 3.34 & 1.31 & 4.06 & 0.172  \\ \hline
        18 & 4.59 & 2.62 & 8.06 & 0.172  \\ \hline
        19 & 13.4 & 5.24 & 16.1 & 0.172  \\ \hline
        20 & 27.8 & 10.5 & 32.1 & 0.172  \\ \hline
    21 & 55.6 & NA & 64.1 & 0.195 \\ \hline
    22 & NA & NA & 1.28e2 & 0.207 \\ \hline
    23 & NA & NA & 2.56e2 & 0.207 \\ \hline
    24 & NA & NA & 5.12e2 & 0.223 \\ \hline
    25 & NA & NA & 1.02e3 & 0.230 \\ \hline
    26 & NA & NA & $>1.5$GB & 0.238 \\ \hline
    27 & NA & NA & $>1.5$GB & 0.254 \\ \hline
    28 & NA & NA & $>1.5$GB & 0.262 \\ \hline
    29 & NA & NA & $>1.5$GB & 0.277 \\ \hline
    30 & NA & NA & $>1.5$GB & 0.297 \\ \hline
    31 & NA & NA & $>1.5$GB & 0.301 \\ \hline
    32 & NA & NA & $>1.5$GB & 0.305 \\ \hline
    33 & NA & NA & $>1.5$GB & 0.320 \\ \hline
    34 & NA & NA & $>1.5$GB & 0.324 \\ \hline
    35 & NA & NA & $>1.5$GB & 0.348 \\ \hline
    36 & NA & NA & $>1.5$GB & 0.352 \\ \hline
    37 & NA & NA & $>1.5$GB & 0.371 \\ \hline
    38 & NA & NA & $>1.5$GB & 0.375 \\ \hline
    39 & NA & NA & $>1.5$GB & 0.395 \\ \hline
    40 & NA & NA & $>1.5$GB & 0.398 \\ \hline
      \end{tabular}
      \\
      (a) & (b)
   \end{tabular}
    \parbox{6in} {
      \parbox{15cm}{\caption{\label{tab:perf1}
        Simulating Grover's algorithm with $n$ qubits
        using Octave (Oct), MATLAB (MAT), Blitz++ (B++) and our simulator QuIDDPro (QP). $>24$hrs indicates that the runtime exceeded our cutoff of 24 hours. $>1.5$GB indicates that the memory usage exceeded our cutoff of 1.5GB. Simulation runs that exceed the memory cutoff can also exceed the time cutoff, though we give memory cutoff precedence. NA indicates that after a cutoff of one week, the memory usage was still steadily growing, preventing a peak memory usage measurement.}} }
  \end{center}
 \end{table}
\begin{table}[!htb]
\small
  \begin{center}
    \begin{tabular}{cc}
      \begin{tabular}{|r|c|c|c|c|}
        \hline
        \multicolumn{5}{|c|}{\bfseries Oracle 2: Runtime (s)}\\
        \hline
        \hline
        $n$ & Oct   & MAT  & B++ & QP \\ \hline
        13    & 1.39e3  & 1.31e2  & 2.47    &  0.617    \\ \hline
        14    & 3.75e3  & 7.26e2  & 5.42    &  0.662    \\ \hline
        15    & 1.11e4    & 4.27e3  & 11.7   &  0.705    \\ \hline
        16    & 3.70e4    & 2.23e4  & 24.9   &  0.756    \\ \hline
    17    & $>24$hrs & $>24$hrs & 53.4 & 0.805 \\ \hline
    18    & $>24$hrs & $>24$hrs & 1.13e2 & 0.863 \\ \hline
    19    & $>24$hrs & $>24$hrs & 2.39e2 & 0.910 \\ \hline
    20    & $>24$hrs & $>24$hrs & 5.15e2 & 0.965 \\ \hline
    21    & $>24$hrs & $>24$hrs & 1.14e3 & 1.03 \\ \hline
    22    & $>24$hrs & $>24$hrs & 2.25e3 & 1.09 \\ \hline
    23    & $>24$hrs & $>24$hrs & 5.21e3 & 1.15 \\ \hline
    24    & $>24$hrs & $>24$hrs & 1.02e4 & 1.21 \\ \hline
    25    & $>24$hrs & $>24$hrs & 2.19e4 & 1.28 \\ \hline
    26    & $>24$hrs & $>24$hrs & $>1.5$GB & 1.35 \\ \hline
    27    & $>24$hrs & $>24$hrs & $>1.5$GB & 1.41 \\ \hline
    28    & $>24$hrs & $>24$hrs & $>1.5$GB & 1.49 \\ \hline
    29    & $>24$hrs & $>24$hrs & $>1.5$GB & 1.55 \\ \hline
    30    & $>24$hrs & $>24$hrs & $>1.5$GB & 1.63 \\ \hline
    31    & $>24$hrs & $>24$hrs & $>1.5$GB & 1.71 \\ \hline
    32    & $>24$hrs & $>24$hrs & $>1.5$GB & 1.78 \\ \hline
    33    & $>24$hrs & $>24$hrs & $>1.5$GB & 1.86 \\ \hline
    34    & $>24$hrs & $>24$hrs & $>1.5$GB & 1.94 \\ \hline
    35    & $>24$hrs & $>24$hrs & $>1.5$GB & 2.03 \\ \hline
    36    & $>24$hrs & $>24$hrs & $>1.5$GB & 2.12 \\ \hline
    37    & $>24$hrs & $>24$hrs & $>1.5$GB & 2.21 \\ \hline
    38    & $>24$hrs & $>24$hrs & $>1.5$GB & 2.29 \\ \hline
    39    & $>24$hrs & $>24$hrs & $>1.5$GB & 2.37 \\ \hline
    40    & $>24$hrs & $>24$hrs & $>1.5$GB & 2.47 \\ \hline
      \end{tabular}
      &
      \begin{tabular}{|r|c|c|c|c|} \hline
        \multicolumn{5}{|c|}{\bfseries Oracle 2: Peak Memory Usage (MB)}\\
        \hline\hline
        $n$ &  Oct   & MAT   & B++ & QP \\ \hline
        13  &  0.218 & 8.22e-2 & 0.252 & 0.137  \\ \hline
        14  &  0.436 & 0.164 & 0.563 & 0.141  \\ \hline
        15  &  0.873 & 0.328 & 1.06 & 0.145  \\ \hline
        16  &  1.74 & 0.656 & 2.06 & 0.172  \\ \hline
    17  &  3.34 & 1.31 & 4.06 & 0.176 \\ \hline
    18  &  4.59 & 2.62 & 8.06 & 0.180 \\ \hline
    19  &  13.4 & 5.24 & 16.1 & 0.180 \\ \hline
    20  &  27.8 & 10.5 & 32.1 & 0.195 \\ \hline
    21  &  55.6 & NA & 64.1 & 0.199 \\ \hline
    22  &  NA & NA & 1.28e2 & 0.207 \\ \hline
    23  &  NA & NA & 2.56e2 & 0.215 \\ \hline
    24  &  NA & NA & 5.12e2 & 0.227 \\ \hline
    25  &  NA & NA & 1.02e3 & 0.238 \\ \hline
    26  &  NA & NA & $>1.5$GB & 0.246 \\ \hline
    27  &  NA & NA & $>1.5$GB & 0.256 \\ \hline
    28  &  NA & NA & $>1.5$GB & 0.266 \\ \hline
    29  &  NA & NA & $>1.5$GB & 0.297 \\ \hline
    30  &  NA & NA & $>1.5$GB & 0.301 \\ \hline
    31  &  NA & NA & $>1.5$GB & 0.305 \\ \hline
    32  &  NA & NA & $>1.5$GB & 0.324 \\ \hline
    33  &  NA & NA & $>1.5$GB & 0.328 \\ \hline
    34  &  NA & NA & $>1.5$GB & 0.348 \\ \hline
    35  &  NA & NA & $>1.5$GB & 0.352 \\ \hline
    36  &  NA & NA & $>1.5$GB & 0.375 \\ \hline
    37  &  NA & NA & $>1.5$GB & 0.375 \\ \hline
    38  &  NA & NA & $>1.5$GB & 0.395 \\ \hline
    39  &  NA & NA & $>1.5$GB & 0.398 \\ \hline
    40  &  NA & NA & $>1.5$GB & 0.408 \\ \hline
     \end{tabular}
     \\
     (a) & (b) \\
    \end{tabular}
    \parbox{6in} {
      \parbox{15cm}{\caption{\label{tab:perf2}
        Simulating Grover's algorithm with $n$ qubits
        using Octave (Oct), MATLAB (MAT), Blitz++ (B++) and our simulator QuIDDPro (QP). $>24$hrs indicates that the runtime exceeded our cutoff of 24 hours. $>1.5$GB indicates that the memory usage exceeded our cutoff of 1.5GB. Simulation runs that exceed the memory cutoff can also exceed the time cutoff, though we give memory cutoff precedence. NA indicates that after a cutoff of one week, the memory usage was still steadily growing, preventing a peak memory usage measurement.}} }
  \end{center}
\end{table}

Before starting simulation of an instance of Grover's algorithm, we
construct the QuIDD representations of Hadamard operators by
incrementally tensoring together one-qubit versions of their matrices
$n-1$ times to get $n$-qubit versions. All other QuIDD operators are
constructed similarly.
Table \ref{tab:growth} shows sizes (in nodes) of respective QuIDDs at
$n$-qubits, where $n=20..100$. We observe that memory usage grows
linearly in $n$, and as a result QuIDD-based simulations of Grover's
algorithm are not memory-limited even at $100$ qubits. Note that this
is consistent with Theorem \ref{theorem:full_unity}.

With the operators constructed, simulation can proceed. Tables
\ref{tab:perf1}a and \ref{tab:perf1}b show performance measurements
for simulating Grover's algorithm with an oracle circuit that searches
for one item out of $2^n$. QuIDDPro achieves asymptotic memory savings
compared to qubit-wise implementations (see Subsection
\ref{sec:matrix_ops}) of Grover's algorithm using Blitz++, a
high-performance numerical linear algebra library for C++
\cite{blitz}, MATLAB, and Octave, a mathematical package similar to
MATLAB. The overall runtimes are still exponential in $n$ because
Grover's algorithm entails an exponential number of iterations, even
on an actual quantum computer \cite{Boyer96}. We also studied a
``mod-$1024$'' oracle circuit that searches for elements whose ten
least significant bits are $1$ (see Tables \ref{tab:perf2}a and
\ref{tab:perf2}b). Results were produced on a 1.2GHz AMD Athlon with
1GB RAM running Linux. Memory usage for MATLAB and Octave is
lower-bounded by the size of the state vector and conditional phase
shift operator; Blitz++ and QuIDDPro memory usage is measured as the
size of the entire program. Simulations using MATLAB and Octave past
15 qubits timed out at 24 hours.

\subsection{Impact of Grover Iterations}

To verify that the QuIDDPro simulation resulted in the exact number of
Grover iterations required to generate the highest probability of
measuring the items being sought as per the Boyer et al. formulation
\cite{Boyer96}, we tracked the probabilities of these items as a
function of the number of iterations. For this experiment, we used
four different oracle circuits, each with $11$,$12$, and $13$ qubit
circuits. The first oracle is called ``Oracle N'' and represents an
oracle in which all the data qubits act as controls to flip the oracle
qubit (this oracle is equivalent to Oracle 1 in the last
subsection). The other oracle circuits are ``Oracle N-1'', ``Oracle
N-2'', and ``Oracle N-3'', which all have the same structure as Oracle
N minus $1$,$2$, and $3$ controls, respectively. As described earlier,
each removal of a control doubles the number of items being searched
for in the database. For example, Oracle N-2 searches for $4$ items in
the data set because it recognizes the bit pattern $111...1dd$.

\begin{table}[!hbt]
\small
  \begin{center}
    \begin{tabular}{|c|c|c|c|} \hline
      Oracle & 11 Qubits & 12 Qubits & 13 Qubits  \\ \hline
      $N$ & 25 & 35 & 50 \\ \hline
      $N-1$ & 17 & 25 & 35 \\ \hline
      $N-2$ & 12 & 17 & 25 \\ \hline
      $N-3$ & 8 & 12 & 17 \\ \hline
    \end{tabular}

    \parbox{4.6in}{
      \caption{
        \label{tab:g_iters}
    Number of Grover iterations at which Boyer et al. \cite{Boyer96} predict
    the highest probability of measuring one of the items sought.
      }
    }
  \end{center}
\end{table}

Table \ref{tab:g_iters} shows the optimal number of iterations
produced with the Boyer et al. formulation for all the instances
tested. Figure \ref{fig:prob} plots the probability of successfully
finding any of the items sought against the number of Grover
iterations.  In the case of Oracle N, we plot the probability of
measuring the single item being searched for. Similarly, for Oracles
N-1, N-2, and N-3, we plot the probability of measuring any one of the
$2$, $4$, and $8$ items being searched for, respectively. By comparing
the results in Table \ref{tab:g_iters} with those in Figure
\ref{fig:prob}, it can be easily verified that QuIDDPro uses the
correct number of iterations at which measurement is most likely to
produce items sought. Also notice that the probabilities, as a
function of the number of iterations, follow a sinusoidal curve. It is
therefore important to terminate at the exact optimal number of
iterations not only from an efficiency standpoint but also to prevent
the probability amplitudes of the items being sought from lowering
back down toward 0.


\begin{figure}[!ht]
  \begin{center}
    \begin{tabular}{cc}
         \includegraphics[width=7.5cm]{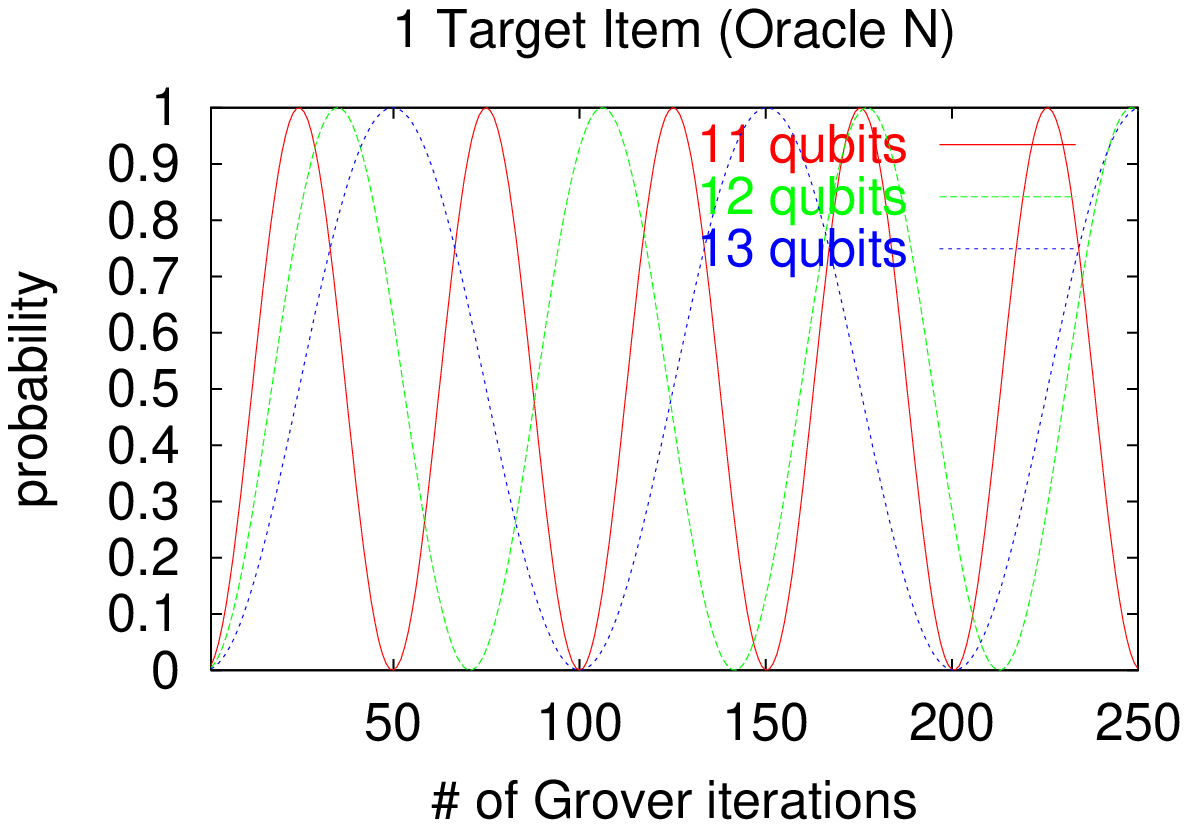} &
         \includegraphics[width=7.5cm]{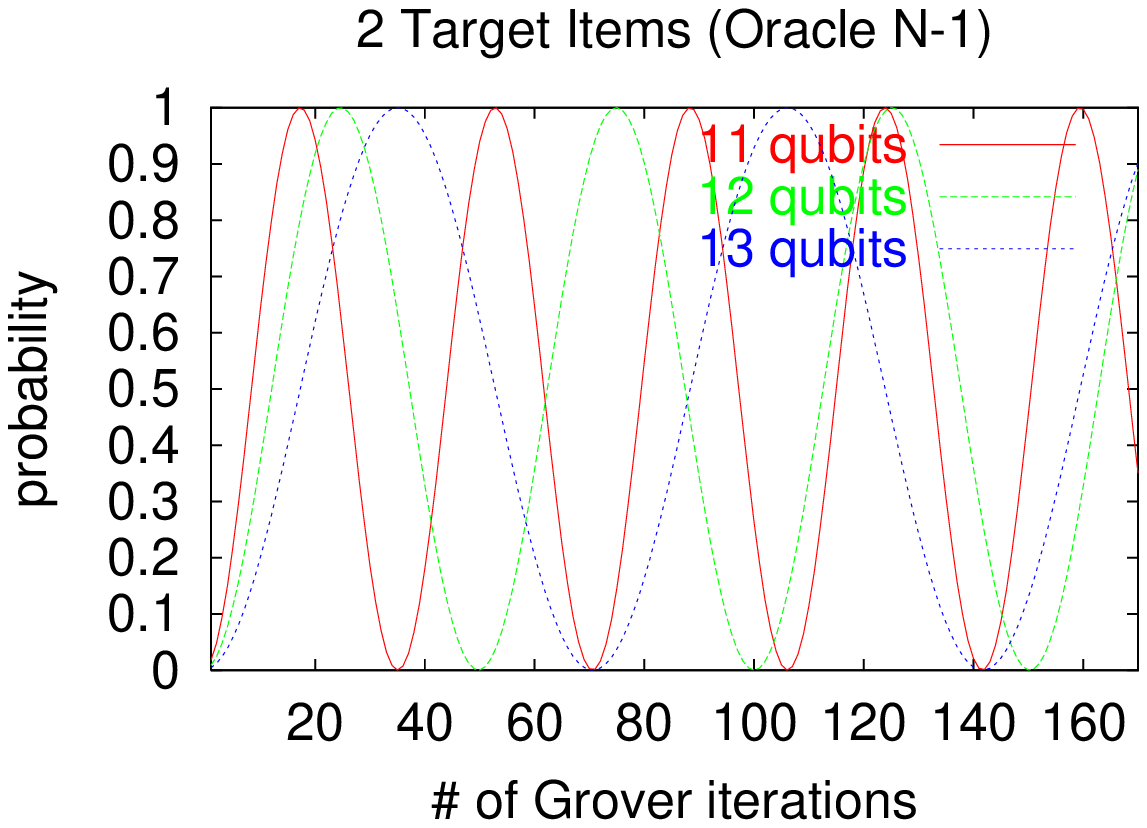} \\
         \includegraphics[width=7.5cm]{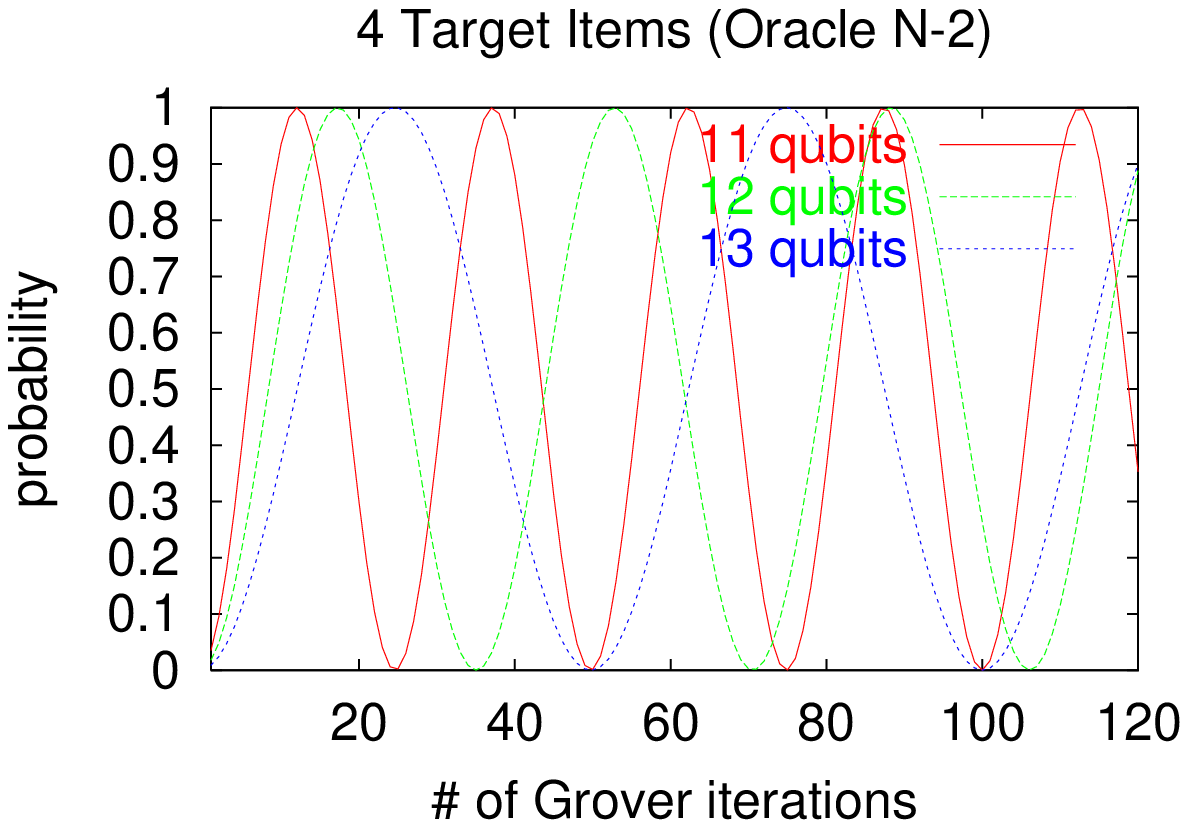} &
         \includegraphics[width=7.5cm]{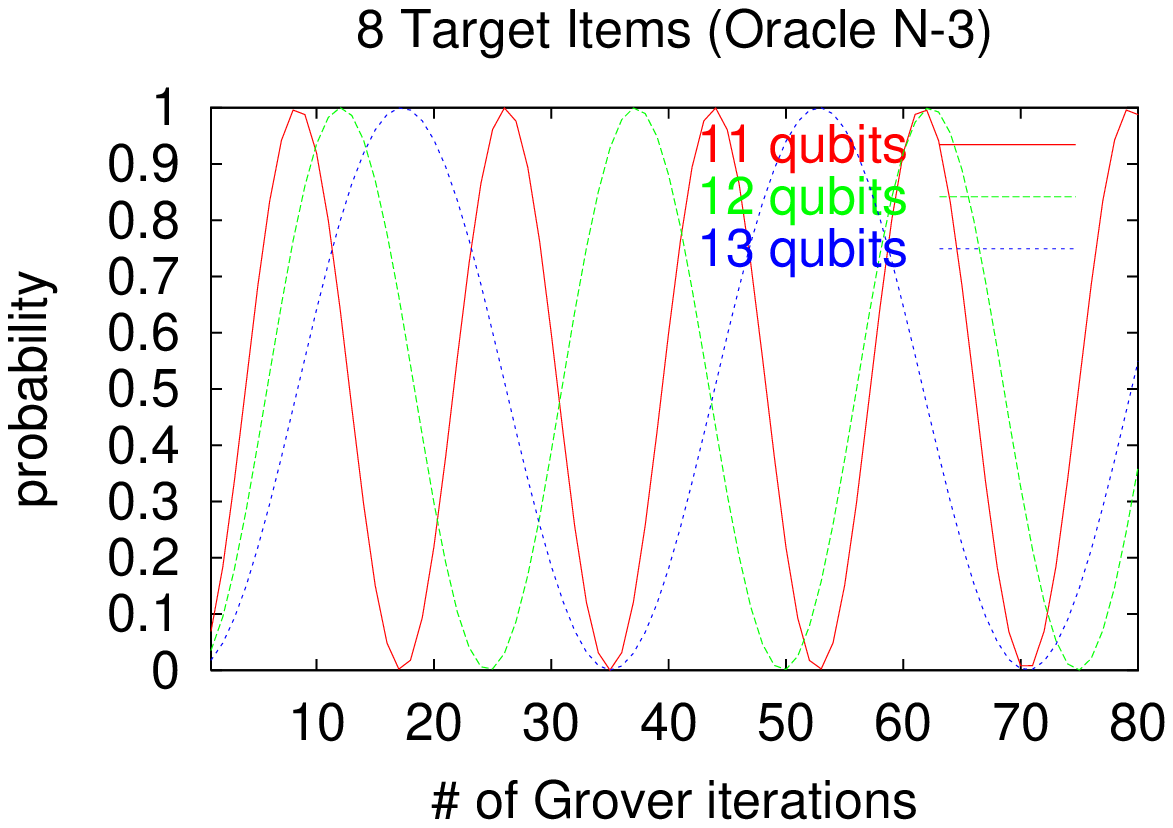}
    \end{tabular}

    \parbox{6in}{
      \caption{
        \label{fig:prob}
        Probability of successful search for one, two, four
        and eight items as a function of the number of iterations
        after which the measurement is performed (11, 12 and 13
        qubits). Note that the minima and maxima of the empirical sine
        curves match the predictions in Table \ref{tab:g_iters}.
      }
    }
  \end{center}
\end{figure}

\section{Conclusions and Future Work}
\label{sec:conclusions}

We proposed and tested a new technique for simulating quantum circuits
using a data structure called a QuIDD. We have shown that QuIDDs
enable practical, generic and reasonably efficient simulation of
quantum computation. Their key advantages are faster execution and
lower memory usage. In our experiments, QuIDDPro achieves exponential
memory savings compared to other known techniques.

This result is a first step in part of our ongoing research which
explores the {\em limitations} of quantum computing. Classical
computers have the advantage that they are not subject to quantum
measurement and errors. Thus, when competing with quantum
computers, classical computers can simply run ideal error-free
quantum algorithms (as we did in Section \ref{sec:exp}), allowing
techniques such as QuIDDs to exploit the symmetries found in ideal
quantum computation. On the other hand, quantum computation still
has certain operators which cannot be represented using only
polynomial resources on a classical computer, even with QuIDDs.
Examples of such operators include the quantum Fourier Transform
and its inverse which are used in Shor's number factoring
algorithm \cite{Shor97}. Figure \ref{fig:inv_qft} shows the growth
in number of nodes of the $N$ by $N$ inverse Fourier Transform as
a QuIDD. Since $N=2^n$ where $n$ is the number of qubits, this
QuIDD exhibits exponential growth with a linear increase in
qubits. Therefore, the Fourier Transform will cause QuIDDPro to
have exponential runtime and memory requirements when simulating
Shor's algorithm.


\begin{figure}
  \begin{center}
    \includegraphics{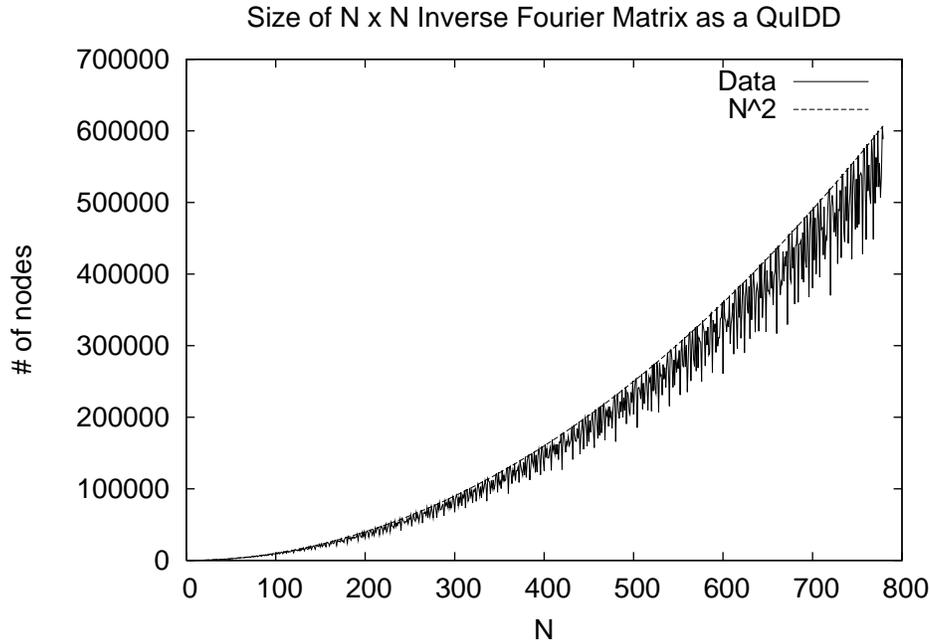}
    \parbox{14cm}
    {
      \caption{ \label{fig:inv_qft}
      Growth of inverse Fourier Transform matrix in QuIDD
      form. $N=2^n$ for $n$ qubits.}
    }
  \end{center}
  \vspace{-4mm}
\end{figure}

Another challenging aspect of quantum simulation that we
are currently studying is the impact of errors due to defects in
circuit components, and environmental effects such as decoherence.
Error simulation appears to be essential for modelling actual
quantum computational devices. It may, however, prove to be
difficult since errors can alter the symmetries exploited by
QuIDDs.

\section*{Appendix: A Characterization of Persistent Sets}

  The following sequence of lemmas leads to a complete
  characterization of persistent sets from Definition \ref{def:persistent}.
  We start by observing that adding 0 to or removing 0 from a set
  does not affect its persistence.

\begin{app_lemma}
        {\em All elements of a persistent set $\Gamma$ that does not contain 0
        must have the same magnitude.}
\end{app_lemma}

{\bf Proof.} In order for $\Gamma$ to be persistent, the set of
        magnitudes of elements from $\Gamma$ must also be persistent.
        Therefore, it suffices to show that each persistent set of
        positive real numbers contains no more than one element.
        Assume, by contradiction, such a persistent set with at least
        two elements $r$ and $s$. Then among $n$-element products from
        $\Gamma$, we find all numbers of the form $r^{n-k}s^k$ for
        $k=0..n$. If we order $r$ and $s$ so that $r<s$, then it
        becomes clear that they are all different because
        $r^{n-k+1}s^{k-1} < r^{n-k}s^k$.\hfill $\Box$

\begin{app_lemma}
\label{lem:normalization}
  {\em All persistent sets without 0 are of the form $c\Gamma'$, where
        $c\neq 0$ and $\Gamma'$ is a finite persistent subset of the
        unit circle in the complex plane $\mathbb{C}$, containing 1 and closed
        under multiplication.  Vice versa, for all such sets $\Gamma'$
        and $c\neq 0$, $c\Gamma'$ is persistent.}
\end{app_lemma}

{\bf Proof.} Take a persistent set $\Gamma$ that does not contain 0,
        pick an element $z\in\Gamma$ and define $\Gamma'=\Gamma/z$,
        which is persistent by construction.  $\Gamma'$ is a subset of
        the unit circle because all numbers in $\Gamma$ have the same
        magnitude.  Due to the fact that $z/z=1\in\Gamma'$, the set of
        $n$-element products contains every element of
        $\Gamma'$. Should the product of two elements of $\Gamma'$
        fall beyond the set, $\Gamma'$ cannot be persistent.  \hfill
        $\Box$

\begin{app_lemma}
  {\em A finite persistent subset $\Gamma'\ni 1$ of the unit circle
       that is closed under multiplication
       must be of the form $\mathbb{U}_n$ (roots of unity of degree $n$).}
\end{app_lemma}

  {\bf Proof.}  If $\Gamma'=\{1\}$, then $n=1$, and we are done.
Otherwise consider an arbitrary element $z\neq 1$ of $\Gamma'$ and
observe that all powers of $z$ must also be in
$\Gamma'$. Since $\Gamma'$ is finite, $z^m=z^k$ for some $m\neq
k$, hence $z^{m-k}=1$, and $z$ is a root of unity. Therefore
$\Gamma'$ is closed under taking the inverse, and forms a
group. It follows from group theory, that a finite subgroup of
$\mathbb{C}$ is necessarily of the form $\mathbb{U}_n$ for some
$n$.$ \hfill \Box$

\begin{app_theorem}
  {\em Persistent sets are either of the form $c\mathbb{U}_n$ for some
        $c\neq 0$ and $n$, or $\{0\}\cup c\mathbb{U}_n$.}
\end{app_theorem}

\end{document}